# Exploiting the dynamics of commodity futures curves

Robert J. Bianchi[a], John Hua Fan[a], Joëlle Miffre[b,c,*], Tingxi Zhang[d]

*a. Griffith Business School, Griffith University, Brisbane, Australia*

*b. Audencia Business School, 8 Route de la Jonelière, 44300, Nantes, France*

*c. Louis Bachelier Fellow, Paris, France*

*d. Curtin University, Perth, Australia*

## Abstract

The Nelson-Siegel framework is employed to model the term structure of commodity futures prices. Exploiting the information embedded in the level, slope and curvature parameters, we develop novel investment strategies that assume short-term continuation of recent parallel, slope or butterfly movements of futures curves. Systematic strategies based on the change in the slope generate significant profits that are unrelated to previously documented risk factors and can survive reasonable transaction costs. Further analysis demonstrates that the profitability of the slope strategy increases with investor sentiment and is in part a compensation for the drawdowns incurred during economic slowdowns. The profitability can also be magnified through timing and persists under alternative specifications of the Nelson-Siegel model.

*Keywords:* Commodity futures; Nelson-Siegel model; Slope strategy; Spread

*JEL classifications:* G12, G13

*This version:* July 16, 2023

[*] Corresponding author, Tel: +33 2 4037 3434.

E-mails: r.bianchi@griffith.edu.au (R. Bianchi), j.fan@griffith.edu.au (J. Fan), jmiffre@audencia.com (J. Miffre) and riven.zhang@curtin.edu.au (T. Zhang).

## 1. Introduction

The model of Nelson and Siegel (1987) (NS hereafter) is widely used to price the term structure of interest rates. The three parameters of the model, interpreted as level, slope and curvature, accurately capture the quasi-flat, upward-sloping, downward-sloping and humped shapes of the zero-coupon yield curve and explain approximately 96% of its observed variation. The popularity of the NS model stems from its parsimony, which induces stability in the estimated parameters, a key feature in the design of successful investment strategies (Diebold and Li, 2006). Following the seminal work of Nelson and Siegel (1987), attempts have been made to price commodity futures curves using the same framework (Heidorn et al., 2015; GrØnborg and Lunde, 2016; Karstanje et al., 2017).[1] Inspired by this research, we employ the NS model to recover commodity futures price curves and then exploit the information embedded in the level, slope and curvature parameters to develop a set of novel investment strategies.

The NS model is found to capture, on average, more than 95% of the variation in commodity futures curves, and thus, it demonstrates outstanding in-sample goodness-of-fit. The success of the NS model stems from both the slope and curvature parameters. Regression estimates reveal that the model captures only 38% (83%) of the variation in the observed price curve in the absence of the slope (curvature) parameter.

We then exploit the information extracted from the level, slope and curvature parameters and design novel long-short investment strategies. The proposed strategies, hereafter labelled as *L* (Level), *S* (Slope) and *C* (Curvature), rely on the notion that recent parallel shifts (*L* strategy), slope movements (*S* strategy) and butterfly oscillations (*C* strategy) of the futures curves will persist one day ahead. The recommended trades are then implemented on a cross section of 21 commodities by taking long and short positions on front-end contracts ($L$ strategy), on slope spreads ($S$ strategy) or on butterfly spreads ($C$ strategy).[2] While the $L$ strategy is unprofitable, the $S$ and $C$ strategies generate large out-of-sample Sharpe ratios equal to 1.41 and 1.27, respectively.

---

[1] Heidorn et al. (2015) employ the NS model to investigate the impact of fundamental and financial traders on crude oil futures markets. GrØnborg and Lunde (2016) use the NS model to forecast oil futures prices. Karstanje et al. (2017) study the comovement across 24 commodity futures using a NS framework.

The remarkable performance of the $S$ strategy survives the consideration of transaction costs and exposure to commodity risk factors based on characteristics such as carry, hedging pressure, curve-momentum, or relative basis, to name only a few. The positive and statistically significant alphas of the $S$ portfolio suggest that the slope strategy we propose differs from the theoretically motivated and economically rational risk premia previously identified in the asset pricing literature. This motivates us to examine its profits further. In particular, we explore the sources of the observed performance confronting two standard explanations: risk versus behavioral mispricing. We note that both justifications have merit in explaining the observed profits. In support of a risk-based (rational) explanation, we show that the $S$ profits in part compensate investors for the drawdowns incurred when economic activity slows down. In favor of a sentiment-based (behavioral) justification, we note that the $S$ profits rise with sentiment. The performance of the $S$ strategy is driven by the short leg which stems from the reluctance of traders to take short positions in high-sentiment periods (Stambaugh et al., 2012). We further corroborate the behavioral explanation and the role of investor psychology by showing that the long (short) leg of the $S$ portfolio performs best on Fridays (on Mondays) when investors are optimistic (pessimistic).

---

[2] In our paper, the term "slope spread" refers to a simultaneous long-short position in front and distant maturity contracts for a given commodity, while the term "butterfly spread" refers to two positions taken in mid-maturity contracts (i.e., the body of the butterfly) alongside two opposite positions taken in the front and distant maturity contracts (i.e., the wings of the butterfly) for a given commodity.

Further tests indicate that investors can benefit from a timing strategy that increases exposure to the *S* strategy as dispersion in the NS slope betas rises. In various robustness checks, we note that the performance of the *S* strategy does not depend upon the variation in test design, such as the use of test assets with longer maturities, the addition of a seasonal adjustment to the NS model, the use of smoothed signals, or the consideration of specific sectors. We observe, however, a weakening of the *S* profits over time which could stem from recent technological innovations that improve market efficiency (Zaremba et al., 2020), from crowding (Kang et al., 2021), or from the prolonged accommodative monetary policy settings of the past decade.

There is a well-established literature that describes the behavior of commodity futures prices via continuous time models with latent state variables.[3] Extensions of this literature to the NS model only arrived recently (Heidorn et al., 2015; GrØnborg and Lunde, 2016; Karstanje et al., 2017). Following these latest endeavors, we design for the first time three NS-based dynamic trading strategies that exploit the information extracted from fitted commodity futures curves. We establish the efficacy of the NS-based slope strategy, and thereby extend the literature on the profitability of commodity strategies based on characteristics such as inventory levels, roll yield, hedging pressure, past performance, skewness, liquidity, basis-momentum and relative basis (Erb and Harvey, 2006; Gorton and Rouwenhorst, 2006; Miffre and Rallis, 2007; Gorton et al., 2013; Yang, 2013; Szymanowska et al., 2014; Fernandez-Perez et al., 2018; Koijen et al., 2018; Bakshi et al., 2019; Boons and Prado, 2019; Gu et al., 2019). As our NS-based slope strategy recommends the trading of spreads instead of outright front-end positions, our paper also contributes to a literature that studies the profitability of spread strategies in commodity futures markets (Szymanowska et al., 2014; Boons and Prado, 2019; Paschke et al., 2020). Finally and to the best of our knowledge, our paper is the first to design a butterfly strategy in commodity futures markets. As such, our work emphasizes the importance of nonlinearities in commodity futures curves where the convexity at the front end arises according to the model of Deaton and Laroque (1992) due to the non-negativity constraint on inventories.

---

[3] Important contributions include Gibson and Schwartz (1990), Schwartz (1997), Casassus and Collin-Dufresne (2005). Heath (2019) recently extended this literature by developing an affine futures pricing model with both unobserved and macroeconomic state variables.

The paper proceeds as follows. Section 2 presents the NS model and details the methodology employed in the design of the NS strategies. Section 3 describes the data and the benchmarks used to appraise performance. Section 4 presents the results pertaining to the three NS strategies, focusing on the in-sample fit of the NS model, transaction costs and out-of-sample risk-adjusted performance. Given the high abnormal returns of the $S$ strategy net of transaction costs, Section 5 focuses on it exclusively, exploring *inter alia* the sources of the observed performance and the potential gains obtained from timing the strategy. Finally, Section 6 presents our conclusion.

## 2. Nelson-Siegel methodology

### 2.1. Nelson-Siegel model

Following Nelson and Siegel (1987), Diebold and Li (2006), and Karstanje et al. (2017), we use the following model to fit the futures curve of a given commodity at time *t*:

$$F_t(M) = \beta_{L,t} + \beta_{S,t}\left(\frac{1-e^{-\lambda_t M}}{\lambda_t M}\right) + \beta_{C,t}\left(\frac{1-e^{-\lambda_t M}}{\lambda_t M} - e^{-\lambda_t M}\right) + \varepsilon_{t,M} \qquad (1)$$

where $F_t(M)$ represents the price of a futures contract with a time-to-maturity of $M$ (expressed in number of months) at time *t*, $\lambda_t$ is the decay factor measured at time *t* and $\varepsilon_{t,M}$ is the error term.[4] $\beta_{L,t}$, $\beta_{S,t}$ and $\beta_{C,t}$ are the level, slope and curvature betas, respectively. For each commodity, the three betas are estimated daily using the nearest four contracts.

The level beta is independent of maturity, affects all futures prices, and hence can be interpreted as the long-term price, i.e., the price when maturity is infinite ($M=+\infty$). The changes in $\beta_{L,t}$ capture parallel shifts in the futures price curve with the curve shifting up when $\Delta\beta_{L,t} > 0$ and down when $\Delta\beta_{L,t} < 0$.

The slope beta measures the sensitivity of futures prices to a function that starts at one and decays to zero as maturity rises. It can be interpreted as the short- to long-term spread in the futures curve, with a negative $\beta_{S,t}$ indicating a contangoed upward-sloping price curve and a positive $\beta_{S,t}$ indicating a backwardated downward-sloping price curve. Changes in $\beta_{S,t}$ capture slope movements in the futures price curve with positive $\Delta\beta_{S,t}$ indicating more backwardation or less contango and negative $\Delta\beta_{S,t}$ suggesting more contango or less backwardation.

The regressor related to the curvature beta is a function that starts at zero, peaks at a certain value, and then diminishes to zero as maturity rises. Therefore, the curvature beta can be viewed as a measure of the concavity (if $\beta_{C,t} > 0$) or convexity (if $\beta_{C,t} < 0$) of the curve. Changes in $\beta_{C,t}$ capture butterfly movements in the futures price curve with positive $\Delta\beta_{C,t}$ denoting more concavity or less convexity and negative $\Delta\beta_{C,t}$ denoting more convexity or less concavity.

---

[4] $\lambda_t$ is the value that maximizes $\left(\frac{1-e^{-\lambda_t M}}{\lambda_t M} - e^{-\lambda_t M}\right)$ when $M$ equals the average maturity of selected contracts along the futures curve at each point in time. There is no definite consensus on the approach to be applied to determine the decay value. Karstanje et al. (2017) apply a constant decay for each commodity, whereas GrØnborg and Lunde (2016) conduct a grid search on intervals between 0.001 and 0.15 and select the value that minimizes the sum of squared error terms over the sample period. To avoid look-ahead bias, this paper opts for a time-varying decay that is estimated daily.

*2.2. Nelson-Siegel strategies*

The information contained in the NS parameters is then exploited to construct three strategies. The strategies rely on the notion that the change in the NS parameters over the previous day is a good predictor of the change in the NS parameters one-day ahead. In keeping with this assumption, we design the following cross-sectional Level (*L*), Slope (*S*) and Curvature (*C*) strategies.

Level (*L*): The *L* strategy sorts the cross section available at the time of portfolio formation in descending order of $\Delta\beta_{L,t}$. It buys the front-month futures contracts of commodities with positive $\Delta\beta_{L,t}$ and shorts those with negative $\Delta\beta_{L,t}$.

Slope (*S*): The *S* strategy sorts the cross section in descending order of $\Delta\beta_{S,t}$ and partitions the data into two portfolios. The first portfolio takes a long slope-spread position (i.e., long the front contract and short the fourth contract) in commodities with positive $\Delta\beta_{S,t}$. The second portfolio takes a short slope-spread position in the remaining commodities (i.e., short the front contract and long the fourth contract). In both settings, we anticipate a steepening of the slope, becoming either more backwardated (long spread) or more contangoed (short spread).

Curvature (*C*): The *C* strategy is deemed to benefit from the butterfly oscillations along the curve. It sorts the cross section in descending order of $\Delta\beta_{C,t}$ and divides it into two portfolios. The first portfolio is a long butterfly-spread strategy (i.e., short the front contract, long the second contract twice and short the fourth contract). This position is deemed to benefit from an increase in the concavity of the curve. The second portfolio is a short butterfly-spread strategy (i.e., long the front contract, short the second contract twice and long the fourth contract). This position is deemed to benefit from an increase in the convexity of the curve. The long butterfly strategy is applied to the commodities with positive $\Delta\beta_{C,t}$ and the short butterfly strategy is implemented on the remaining markets.

Irrespective of the strategy considered, the NS model is estimated daily. The strategies are set up at the end of day $t$ and implemented for one day. They are fully collateralized and allocate an equal amount of capital to the longs and shorts; thus, the weights assigned to the individual contracts are not necessarily the same. Taking the $L$ strategy as an example, this translates into $\sum_{j=1}^{l} w_{j,t} = \sum_{k=1}^{N-l} |w_{k,t}| = 0.5$ where $w_{j,t}$ is the weight assigned to contract $j$ at time $t$, $l$ is the number of contracts with $\Delta\beta_{L,t} > 0$ and $N$ is the total number of contracts traded at portfolio formation time $t$. Likewise, the $S$ ($C$) strategies allocate an equal amount of capital to the long and short slope (butterfly) spreads, again assuming full collateralization.

## 3. Data and performance benchmarks

### 3.1. Futures contracts

Our sample consists of 21 commodity futures from seven sectors: Energy (crude oil, gasoline, heating oil), Grains (corn, oats, rough rice, wheat), Industrial materials (cotton, lumber), Meats (feeder cattle, live cattle, live hogs), Metals (copper, gold, silver), Oilseeds (soybean meal, soybean oil, soybeans) and Softs (cocoa, coffee, orange juice). The cross section of markets follows Szymanowska et al. (2014). Daily settlement price, volume and open interest are obtained from Bloomberg and the Commodity Research Bureau (CRB), while weekly trader position data are retrieved from the Commodity Futures Trading Commission (CFTC) commitments of traders (CoT) report. The sample covers the period from January 1992 to June 2019 with the starting date dictated by data availability of the CoT report. To compile the time-series variables (i.e., return, settlement price, volume, and open interest), we assume that investors hold a given contract until the last trading day before the front contract enters its maturity month. One-period excess return is defined as $r_t = \frac{F_t}{F_{t-1}} - 1$, where $F_t$ denotes the settlement price for a given contract at day $t$.

Table 1 presents the annualized mean and annualized standard deviation of the excess returns of long outright positions in the front to fourth contracts, long slope spread positions (long in the front contract and short in the fourth contract) and long butterfly spread positions (short in the front contract, long in the second contract twice and short in the fourth contract). All positions are fully collateralized. Figure 1 illustrates the open interest of the front to 12[th] contracts as averaged over time and across commodities.

[Insert Table 1 and Figure 1 around here]

With only a few exceptions, the excess returns of long outright positions are not statistically different from zero, confirming the absence of a long-only risk premium at the individual commodity level (Erb and Harvey, 2006). Volatility is high especially at the front end of the curve. Liquidity decreases along the futures curve (Figure 1), with the nearest four contracts representing on average 82.2% of total open interest (as shown in the last column of Table 1 and in Figure 1). These liquidity empirics support our choice of maturities.

Similar to the long outright positions, most long slope spreads and long butterfly spreads exhibit insignificant mean excess returns at the 10% significance level. We note some propensity for the futures curves to be upward-sloping (5 long slope spread strategies exhibit negative mean excess returns at the 10% level, suggesting that a short slope spread is profitable) and convex (7 long butterfly strategies yield negative mean excess returns at the 10% level, suggesting that a short butterfly strategy is optimal). It is noticeable that the excess returns of both the long slope and long butterfly strategies are less volatile than those of long outright positions, which highlights the hedging effectiveness of spreads (Melamed, 1981).

*3.2. Benchmarks*

To evaluate the performance of the *L, S* and *C* strategies, two types of commodity benchmarks are constructed: naive and traditional risk factors. Table 2 presents summary statistics of their performance.[5] Similar to the NS strategies, the naive benchmark LAVG is an equal-weight, long-only, market-wide level portfolio based on front contracts, while the SAVG (CAVG) is an equal-weight, long-only portfolio of slope (butterfly) spreads. All three naive benchmark portfolios are rebalanced daily. SAVG and CAVG incur yearly average losses of 0.46% and 0.27%, respectively, with the latter being significant at the 5% level. The average LAVG return is indistinguishable from zero. This result is consistent with the lack of statistical significance observed in Table 1 for the long front positions ($M = 1$), the long slope spreads, and the long butterfly spreads as implemented at the individual commodity level.

[Insert Table 2 around here]

Moreover, nine traditional risk factors are constructed, namely, a long-only market portfolio (AVG), momentum (MOM), carry (CARRY), hedging pressure (HP), skewness (SKEW), basis-momentum (BMOM), relative basis (RB), liquidity (LIQ) and curve momentum (Curve-M). The sorting signals and representative studies are detailed in Appendix A. The constituents are equal-weighted, while the positions are fully collateralized and rebalanced at the end of each month. In keeping with the NS strategies, 50% of the exposure is invested in the longs and likewise in the shorts.[6] As reported in Table 2, Panel B, all portfolios except for AVG and LIQ exhibit positive mean excess returns at the 10% level.

Following the recent factor zoo literature, we test whether the excess returns of the NS strategies reflect compensation for exposure to risk factors that are not specific to commodity futures markets. Similar to Ilmanen et al. (2021), we consider the long-short factors based on value, momentum and carry across all asset classes, as well as data on sentiment, macroeconomic and financial variables. Section 5 provides further information on these variables.

---

[5] All Newey and West (1987) adjusted *t*-statistics in this study employ the Bartlett kernel to estimate the lag length truncation parameter.

[6] It is worth noting that our estimated traditional risk premia are generally lower than previously documented in the literature for three reasons. First, the long and short portfolios contain 50% of the available cross section (rather than the extreme quintiles as typically considered). As a result, the signals may be less strong. Second, we assume no leverage, while others (e.g., Miffre and Rallis, 2007) assume a leverage of 2. Third, as argued by Kang et al. (2021), a crowding effect has led to a deterioration in factor mean returns in the most recent period.

## 4. Main empirical results

### 4.1. In-sample goodness-of-fit

The novelty of employing the NS model in commodity futures is that it widens the perspective beyond the front contract in understanding futures price dynamics. This leads to the question of how capable the NS model is at capturing the shape of the commodity futures curve. To investigate this issue, we estimate Equation (1) daily from January 1992 to June 2019 for each commodity curve. Table 3 presents summary statistics of the estimated NS parameters and goodness-of-fit statistics. Figure 2 plots the level betas, slope betas and settlement prices of the front contracts of selected commodities.

[Insert Table 3 and Figure 2 around here]

The column labelled "Average $R^2$" highlights the excellent fit of the NS model as it explains, on average, 96.5% of the in-sample variation in commodity futures curves. The last two columns document the impact on the goodness-of-fit statistics of omitting either the slope or curvature components from Equation (1). The model fit decreases to 37.5% and 82.7%, on average, when the slope or curvature component is omitted, respectively. Clearly, all three components are essential in fitting commodity futures curves, suggesting that the term structures of these markets are subject to both parallel and non-parallel shifts over time.

The average level betas are always positive, which is expected since $\beta_L$ measures long-term settlement prices. Out of the 21 average slope betas, 17 are negative (at the third decimal), suggesting that commodity markets were often in contango over the period from January 1992 to June 2019. Figure 2 confirms the positivity of the level betas and the tendency for the slope betas to be negative, for example, for gold and corn. Because the average curvature betas are either positive or negative, we do not note a propensity for the futures curves to be either concave or convex. It is noticeable, however, that all the commodities with convex curves (or with negative average curvature betas) happen to also be, on average, in contango (i.e., they exhibit negative average slope betas). This figuratively translates into a steep rise of the distant part of the curve in conjunction with a front-end that is relatively flattened. Lastly, even though some commodities report high standard deviations in Table 3, one should not misinterpret this as a lack of persistence in the estimated betas over time. The daily changes in beta indeed exhibit strong persistence and the majority exhibits *p*-values indicating a rejection of no serial correlation.

### 4.2. NS strategies

The excellent fit of the NS model leads us to the next discussion. Can we exploit the information embedded in the level, slope and curvature parameters to develop a set of novel investment strategies that are profitable out-of-sample? To answer this question, Table 4 presents summary statistics for the performance of the long, short and long-short cross-sectional portfolios based on the *L*, *S* and *C* strategies, respectively.[7]

[Insert Table 4 around here]

Over the period spanning from January 1992 to June 2019, the *L* strategy fails to deliver a significantly positive mean excess return, suggesting that the prices of the front contracts that are traded substantially diverge from their long-run levels measured by $\beta_L$. This result is not surprising given our previous conclusion regarding the importance of modeling the slope and curvature of the futures curves (Table 3). In sharp contrast, the *S* strategy based on the change in the slope beta delivers an annualized mean excess return of 1.77% that is highly significant (*t*-statistic=7.23) and a Sharpe ratio of 1.41. Likewise, the *C* strategy based on the change in the curvature beta generates statistically significant profits, with a mean excess return of 0.68% a year, a *t*-statistic of 5.34 and a Sharpe ratio of 1.23. The annualized standard deviation of the *C* strategy equals 0.55% versus 1.25% for the *S* strategy and 6.30% for the *L* strategy. Consistent with the results presented in Table 1, trading butterfly spreads is the least risky investment strategy. It is worth noting that the performance of the long-short *S* and *C* strategies is mostly driven by the underperformance of the short legs. We will discuss this result in Section 5.2 when analyzing the impact of investor sentiment on performance.

---

[7] The CER (power utility certainty-equivalent-return) is calculated as $CER = \left(\frac{n}{T}\right)\sum_{t=0}^{T-1}\frac{\left(1+r_{P,t+1}\right)^{1-\gamma}-1}{1-\gamma}$, where $r_{P,t+1}$ represents the strategy portfolio return at time $t+1$, $n$ is the total number of trading days in a year and $\gamma$, the relative risk aversion parameter, is set to 5. A positive CER indicates the strategy is superior to the risk-free asset.

The success of the *S* and *C* strategies indicates that a compelling investment opportunity arises from harnessing the information embedded in the slope and curvature betas. While the slope and curvature have been used as trading signals before (Erb and Harvey, 2006; Gorton and Rouwenhorst, 2006; Gu et al., 2019), we are the first to use the NS model to extract the slope and curvature of futures curves and to show that they hold profitable information out-of-sample.

It is worth noting that the absolute annual excess returns of the *S* and *C* strategies are statistically significant, but small. This is due to the low-risk profiles of slope and butterfly spreads (Table 1) and to the fully-collateralized nature of our long-short portfolios. In practice, the excess returns of these investment strategies can be magnified through leverage. This is easy to manage in futures markets given that the required margins are much lower than the notional values of the contracts. This is even more true for spread positions as the futures margins can be as much as 80% lower than those required on outright positions (Dunis et al., 2006).

Motivated by Moskowitz et al. (2012), we also implement the *L*, *S* and *C* cross-sectional strategies in a time-series setting. Taking the *L* strategy as an example, we measure the $\Delta\beta_{L,i,t}$ signal of each commodity curve *i* ($i = 1,\ldots, N$) at the end of day *t*, we take a long position in the front-end contract of the commodity *i* if $\Delta\beta_{L,i,t} > 0$ and a short position if $\Delta\beta_{L,i,t} < 0$. Repeating the same strategy for all commodities ($i = 1,\ldots, N$), we subsequently form an equal-weighted and fully-collateralized portfolio that we hold for a day. Unlike the cross-sectional portfolio, the time-series portfolio therefore takes an equal position in each commodity. We proceed likewise with the $\Delta\beta_{S,i,t}$ and $\Delta\beta_{C,i,t}$ signals of the *S* and *C* time-series strategies, with the only difference being that we then trade either slope spreads or butterfly spreads in place of the front contracts.[8] The time-series results are very similar to the cross-sectional results reported in Table 4. For brevity, we omit these results, but they are available upon request.

### 4.3. Turnover and transaction costs

Our NS strategies are trading intensive since they assume daily rebalancing. It is therefore important to evaluate the impact of transaction costs on performance. We center our attention on the long-short strategies with positive and statistically significant mean excess returns in Tables 2 and 4 and calculate their turnover (TO) as follows:

$$TO = \frac{1}{T-1}\sum_{t=1}^{T-1}\sum_{i=1}^{N}\left(\left|w_{t+1}^{c,i} - w_{t^+}^{c,i}\right|\right) \tag{2}$$

where $w_{t+1}^{c,i}$ is the weight of contract *i* for commodity *c* at day *t*+1 as dictated by the fully-collateralized strategy under consideration, $w_{t^+}^{c,i}$ denotes the actual weight of contract *i* for commodity *c* at the end of day *t*+1 prior to the rebalancing of the strategy and after accounting for the performance of the contract from $t$ to $t+1$; namely, $w_{t^+}^{c,i} = w_t^{c,i}(1 + r_{i,t+1})$ where $r_{i,t+1}$ is the excess return of contract $i$ from day $t$ to day $t+1$ and $N$ is the total number of contracts traded at time $t$ under a given strategy. The turnover is a measure of trading intensity that ranges from 0 (when there is no trading) to 2 (when all positions are reversed). It is worth emphasizing that our turnover estimates account for roll-over effects[9] and the sum of the absolute weights of all contracts involved at each time $t$ equals 100% (to be compatible with full collateralization).

---

[8] These assumptions translate into portfolio weights: $\left|w_{i,1,t}\right| = \frac{1}{N}$ for the *L* strategy; $\left|w_{i,1,t}\right| = \left|w_{i,4,t}\right| = \frac{1}{2N}$ for the *S* strategy and $\left|w_{i,1,t}\right| = \left|w_{i,4,t}\right| = \frac{1}{4N}$ and $\left|w_{i,2,t}\right| = \frac{1}{2N}$ for the *C* strategy, where *j* in $w_{i,j,t}$ designates the 1st, 2nd or 4th contract as present in the term structure of commodity *i* at time *t*.

Net returns, $\tilde{r}_{P,t}$, are measured as:

$$\tilde{r}_{P,t} = r_{P,t} - 0.5 \times \sum_{i=1}^{N} TO_t^{c,i} \times TC_t^{c,i} \tag{3}$$

where $r_{P,t}$ is the gross return at time $t$ as previously calculated, $TO_t^{c,i} = \left| w_t^{c,i} - w_{t-1^+}^{c,i} \right|$ and $TC_t^{c,i}$ denote the turnover and transaction cost for contract $i$ of commodity $c$ at time $t$, respectively.[10] To keep transaction costs as low as possible, we assume that investors who implement the $S$ strategy do not execute outright positions in two contracts but rather directly trade spreads and likewise, investors who follow the $C$ strategy trade two spreads as opposed to four contracts.[11]

We apply the following three transaction cost scenarios:

---

[9] When rolling contracts, the turnover at time $t$ is calculated as $\left| w_{t-1^+}^{c,i_1} \right| + \left| w_t^{c,i_2} \right|$ where $i_1$ pertains to the contract that is closed out and $i_2$ pertains to the contract that is entered into.

[10] Following Paschke et al. (2020), we multiply the transaction cost by 0.5 in Equation (3) to account for round-trip costs.

[11] Market participants can trade futures spreads as any multi-legged instrument comprised of outright futures and/or futures spreads. As opposed to composite spreads which requires the trader to leg the two contracts separately, listed spreads only cross the bid-ask spread once and requires much lower margins. For example, the below table shows exchange-recognized spread types on CME Globex: https://www.cmegroup.com/confluence/display/EPICSANDBOX/Spreads+and+Combinations+Available+on+CME+Globex

$$\text{Scenario 1: } TC1_t^{c,i} = \frac{C}{F_t^{c,i} \times M_c}$$

$$\text{Scenario 2: } TC2_t^{c,i} = 0.0167\%$$

$$\text{Scenario 3: } TC3_t^{c,i} = \frac{C + n \times Tick_c \times M_c}{F_t^{c,i} \times M_c}$$

$C$ is the commission fee per contract ($C = 1.5$ as suggested by Gao et al., 2018), $M_c$, $Tick_c$ and $F_t^{c,i}$ are the multiplier, tick size and settlement price of contract $i$ for commodity $c$ at time $t$, respectively, and $n$ is the number of tick size assumed in estimating price impact. Following Szakmary et al. (2010) and Paschke et al. (2020), TC1 and TC3 measure transaction costs as the percentage of the sum of the commission fee and/or price impact to total contract value, while TC2 is the average transaction cost of commodity futures documented in Locke and Venkatesh (1997). In terms of price impact, recall that all strategies in this study transact at the end of day settlement price. As a conservative measure, we introduce an additional price impact component, computed as a quarter of tick size ($n = 1/4$ in Scenario 3), as in Locke and Venkatesh (1997).

The results are summarized in Table 5 for the strategies with significant mean excess returns in Tables 2 and 4. The daily turnover of the $S$ ($C$) strategy equals 1.17 (1.20) which signifies that the $S$ and $C$ strategies are trading intensive. The net mean excess returns of the $S$ strategy range from 0.55% to 1.31% per year and are statistically significant at the 5% level or better. The corresponding Sharpe ratios remain high with a range from 0.44 to 1.04. They exceed, or are equal to, the net Sharpe ratios of the traditional risk factors depending on the transaction cost scenario considered.[12] We note however that the net profits of the $C$ strategy are much lower with a range from -0.55% to 0.24% per year and are statistically less than zero under both TC2 and TC3. This transaction cost analysis demonstrates that investors should proceed with caution when implementing the proposed *S* and *C* strategies. Only institutions that can access commodity futures markets at low transaction costs can take advantage of temporal mispricings in commodity futures curves. Given that each institution is subject to different levels of transaction costs depending on their size and sophistication, practitioners may wish to further calibrate the NS model and backtest procedures to suit their specific trading needs.

---

[12] Despite their relatively high turnover (at 1.36, on average), the traditional strategies are found to exhibit positive net mean excess returns under all three TC scenarios. This is due to their relatively less frequent (end-of-month) rebalancing.

[Insert Table 5 around here]

*4.4. Are the excess returns of the S and C strategies related to risk premia?*

Are the *S* and *C* profits merely compensation for exposure to premia obtained from naive benchmarks and classical asset pricing theories? Or do they represent compensation for the skills of selecting mispriced contracts along the futures curves? We answer these questions through time-series regressions of the excess returns of the *S* and *C* strategies on various risk premia. Table 6 presents the coefficient estimates of these spanning regressions, with corresponding Newey and West (1987) adjusted *t*-statistics in parentheses. Panel A treats the excess returns of the naive long-only level, slope and curvature portfolios as independent variables, Panel B considers Bakshi et al. (2019) factors as well as additional characteristic-sorted risk factors and finally, Panel C tests the hypothesis that the observed excess returns represent compensation for global risks that do not solely originate from commodity markets. Bearing this in mind, we consider risk factors including the excess returns of long-short portfolios based on value, momentum and carry across all asset classes, as these factors are deemed to capture funding liquidity, global recessions, and volatility risks (Asness et al. 2013; Koijen et al., 2018). To allow comparison with the mean excess returns presented in Table 4, the intercepts (alphas) are annualized. Since the NS strategies are more sensitive to transaction costs than conventional commodity strategies, we report regression results based on net returns (TC1).

The slope coefficients of the regressions are, for the most part, insignificant at the 5% level. Interestingly, Table 6, Panel B shows that the $S$ and $C$ strategies exhibit insignificant loadings with regards to the carry, curve-momentum, and relative basis strategies, respectively, and thus, the NS-based slope and curvature strategies proposed here are different from the slope, spread and curvature strategies previously documented in the literature. The alphas of the $C$ strategy are at times indistinguishable from zero, suggesting that once we account for risk and transaction costs, the curvature signal is priced efficiently in the term structure of futures prices. However, the net alphas of the $S$ strategy are statistically significant at the 1% level and equal to 1.24% a year, on average, which serves to justify our focus on the $S$ strategy hereafter.[13]

## 5. Further investigation of the performance of the slope strategy

Our results thus far indicate strong predictability of commodity futures returns based on expected changes in the slope parameter. This section first explores the sources of the observed performance challenging two alternative explanations: risk versus sentiment-based mispricing. We also test *inter alia* whether the $S$ profits can be magnified through timing and whether they persist over time, across sectors or under alternative specifications of the NS model. We conclude with a performance analysis of slope strategies that do not rely on the NS model.

---

[13] In unreported results, we consider various equity, fixed income, and currency risk factors emanating from the factor zoo literature and confirm that the abnormal $S$ profits are robust to these alternative factor models. The risk premia considered emanate from i) equity markets (via the five long-short portfolios of Fama and French, 2015 and the momentum portfolio of Carhart, 1997), ii) fixed-income markets (via the market, value, momentum, carry and defensive portfolios of Ilmanen et al., 2021), and iii) currency markets (via the value, momentum and carry portfolios of Ilmanen et al., 2021). The data observations are from the websites of Prof. K. French and AQR.

## 5.1. Risk-based explanation

If the risk-based explanation holds, then the *S* profits reported in Table 4 may be related to macroeconomic and financial risk factors in such a way that the premium earned in "good" times (e.g., when economic activity improves) compensates investors for the drawdowns incurred in "bad" times (e.g., when financial conditions worsen). Following Heath (2019) or Ilmanen et al. (2021), we study the explanatory power of the following macroeconomic and financial risk factors: *i*) term spread (TERM), *ii*) TED spread (TED), *iii*) default spread (DEF), *iv*) the change in the FED funds rate ($\Delta$FED), *v*) the 3-month real interest rate (REAL), *vi*) innovations in aggregate liquidity of Pastor and Stambaugh (2003) (LIQUID), *vii*) the change in industrial production ($\Delta$IP), viii) the Chicago Fed National Activity Index (CFNAI), *viii*) inflation (INFL), and *ix*) the Uncertainty Index (UNC) of Bekaert et al. (2022).[14]

Table 7, Panel A reports the slope coefficients obtained from a contemporaneous time-series regression of the long-short *S* excess returns net of transaction costs onto these risk factors over the period from January 1992 to June 2019. We note a propensity for the long-short *S* profits to be positively related with TERM and REAL. This finding, consistent with rational pricing, indicates that the risk premium identified in Table 4 is earned mainly in periods of economic expansion (when the yield curve slopes upward, and interest rates are relatively

---

[14] TERM is the yield difference between 10-year T-bonds and 3-month T-bills. TED is the difference between 3-month U.S. LIBOR rate and 3-month U.S. T-bill rate. DEF is the yield difference between Moody's seasoned Baa corporate bonds and 10-year constant-maturity T-bonds. REAL is the difference between the 3-month US T-bill rate and expected inflation (3-year moving average of year-on-year change in US consumer price index). $\Delta$IP is the global year-on-year change in industrial production measured as an equal-weighted average of the U.S., U.K., Japan, and Eurozone series. INFL is the year-on-year percentage change in G7 consumer price index (all items). All variables have been tested to ensure stationarity in the time series. The observations are from Refinitiv Datastream and from the websites of the Federal Reserve Bank of St Louis, Prof. R. Stambaugh and Prof. N. Xu.

higher) and that it represents compensation for the drawdowns incurred during economic slowdowns (when the yield curve slopes downward, and interest rates are relatively lower).[15]

[Insert Table 7 around here]

## 5.2. Sentiment-based mispricing

It has long been argued in the literature that sentiment-driven investors may cause prices to deviate from their fundamental values. If rational traders fail to arbitrage away these sentiment effects, then mispricing may persist and may explain the $S$ profits observed in Tables 4 and 6. To test this consideration, we follow Stambaugh et al. (2012) by regressing the excess returns of the $S$ strategy net of transaction costs onto the orthogonalized sentiment index of Baker and Wurgler (2006).[16] We also measure the mean excess returns of the $S$ strategy, as well as its alpha (relative to the three-factor model of Bakshi et al., 2019), in high and low sentiment periods, where the periods are defined relative to the sentiment index full-sample average.

Table 7, Panel B shows that the performance of the long-short $S$ portfolio improves with sentiment. For example, the results of the time-series regression indicate a significant positive relationship between the long-short $S$ profits and investor sentiment ($\beta(BW)$=0.0011, $t$-statistic of 2.56). We also note that the mean excess return of the long-short $S$ portfolio is

---

[15] In unreported results, we considered the macro uncertainty index (TMU) of Jurado et al. (2015) in place of UNC and found consistent results. Furthermore, we note that, while the $S$ strategy outperforms in high REAL episodes, the REAL variable is poor at explaining the variation in the $S$ returns beyond the other macro factors considered (i.e., 2-3% increase in adjusted- $R^2$s).

[16] The choice of the orthogonalized index is governed by the fact that it is presumably free from considerations relating to macroeconomic and financial conditions and thus, it is a cleaner measure of investor psychology than the original index. Unreported results indicate that the conclusion on the presence of a sentiment effect holds irrespective of the sentiment index considered and is unaltered by the consideration of the one-month lagged, in place of the contemporaneous, value of the sentiment index. The observations for the sentiment indexes are from the website of Prof. J. Wurgler.

stronger during high than in low sentiment periods, with the difference being statistically significant (*t*-statistic of 3.58). Likewise, the alpha of the long-short *S* portfolio in high sentiment periods (at 2.61%) is higher than the alpha in low sentiment periods (at 0.5%), with the difference being statistically significant (*p*-value of 0.00). Overall, optimistic sentiment is a key driver of the long-short profits of the *S* portfolio.

Following Stambaugh et al. (2012), we also analyze the performance of the short leg of the *S* portfolio in periods of high versus low sentiment. The results indicate lower excess returns of the short leg when sentiment is high. For example, the relation between the short-leg excess returns and sentiment is negative ($\beta(BW)$=-0.0012, *t*-statistic of -1.86) and the constituents of the short leg are more overpriced (at -3.24%) when sentiment is high than when sentiment is low (at -0.61%) with the difference in mean excess returns in high versus low sentiment periods that is statistically significant (*t*-statistic of -3.60). Similar inference can be drawn from the high versus low sentiment alphas. This suggests that the sentiment effect observed above for the long-short *S* portfolio reflects the reluctance of traders to take short positions (due to limited knowledge or behavioral biases), in periods when assets tend to appreciate; namely, in high sentiment periods. As a result, the spreads included in the short leg of the *S* portfolio tend to be particularly overpriced. Finally, we note mixed evidence regarding the impact of investor sentiment on the profits of the long leg. Indeed, the results show that the profits in the long leg are positively related with sentiment ($\beta(BW)$=0.0011, *t*-statistic of 2.27) but we note no statistical difference between the mean excess returns (alphas) of the long portfolio in high versus low sentiment periods.

These conclusions are consistent with our earlier results of Table 4 that show the performance of the long-short *S* portfolio is more driven by the overpricing of the spreads in the short leg than by the underpricing of the spreads in the long leg. Taking this evidence together, it appears that sentiment and the reluctance of traders to take short positions drive the overpricing of the short leg and provide a partial explanation for the observed profits of the *S* portfolio.

We further explore the role of sentiment by relating the profits of the *S* strategy to investor psychology. For example, Birru (2018) points towards investors' pessimism on Mondays and optimism on Fridays. If sentiment and investor psychology play a role in commodity futures, one expects worse performance of the long leg on Mondays and better performance on Fridays and vice versa for the short leg. Table 7, Panel C presents these results which corroborate this hypothesis. For example, the performance of the long leg net of transaction costs on Mondays (at -3.02% per year, *t*-statistic of -3.37) is more than 7 percentage points lower than that obtained on Fridays (at 4.12% a year, *t*-statistic of 4.88); and vice versa, the losses on Mondays on the short leg (at -4.93% a year, *t*-statistic of -6.11) are over 6 percentage points lower than the gains on Fridays (at 1.43% a year, *t*-statistic of 2.01). This result validates the role of investor psychology as an explanation for the *S* profits. Taken together, Sections 5.1 and 5.2 demonstrate that the performance of the *S* strategy derives from both risk-based and behavioral determinants.

### 5.3. Timing the slope factor

This section employs factor timing to capture the predictable variation (if any) in the *S* factor as an investment strategy. If the NS slope betas ($\beta_{S,t}$ in Equation (1)) are indicative of time-varying excess returns, we can expect a positive relationship between the dispersion in the NS slope betas and forthcoming portfolio returns. As an initial evaluation of the potential gains from timing, we estimate the following regression:

$$r_{S,t+1} = a + bSD\left(\beta_{S,t}\right) + e_{t+1} \qquad (4)$$

where $r_{S,t+1}$ is the excess return of the long-short *S* strategy at time $t+1$, $SD\left(\beta_{S,t}\right)$ is the cross-sectional standard deviation of $\beta_{S,t}$ at time $t$, and $e_{t+1}$ is an error term. Over the period from January 1992 to June 2019, the *t*-statistic for the hypothesis $b = 0$ in Equation (4) equals 3.37. Thus, the wider the dispersion in the estimated NS slopes, the better the performance of the *S* strategy one day ahead.

Bearing this result in mind, we time the allocation to the *S* strategy in such a way that exposure increases with the dispersion in the NS slopes. More precisely, the excess return of the timed *S* portfolio at the end of day *t*+1, $r_{S,t+1}^{\sigma}$, is calculated as:

$$r_{S,t+1}^{\sigma} = \frac{\hat{\sigma}_t}{c} r_{S,t+1} \tag{5}$$

where $\hat{\sigma}_t = \frac{1}{d}\sum_{j=0}^{d-1} SD\left(\beta_{S,t-j}\right)$ is the cross-sectional standard deviation in the estimated NS slopes as averaged over the $d = \{3, 5, 10, 15, 22\}$ days preceding the timing decision taken at day *t*, *c* is a scaling factor that ensures that the original and timed *S* strategies have the same volatility, and $r_{S,t+1}$ is the excess return of the original *S* portfolio at the end of day *t*+1. Essentially, Equation (5) posits that wider spreads in the NS slopes across commodities as estimated over the *d* days preceding time *t* indicate better performance of the *S* strategy expected at time *t*+1 and thus, higher exposure to the *S* strategy at time *t* (namely, $\frac{\hat{\sigma}_t}{c} > 1$). Vice versa, smaller spreads in the NS slopes indicate lower expected performance of, and thus lower exposure to, the *S* strategy (i.e., $\frac{\hat{\sigma}_t}{c} < 1$).

The left Y-axis of Figure 3 plots on the left-hand side the future value of \$1 invested in the original *S* strategy and its timed counterpart (*d*=5) over the period January 1992 to June 2019. The right Y-axis on the right-hand side depicts the leverage of the timed strategy. Given the volatility target, the timed *S* strategy does not require excessive leverage (max. < 2.5x). Table 8 reports summary statistics for the timed *S* strategy, as well as estimated parameters of a spanning regression of $r_{S,t}^{\sigma}$ onto $r_{S,t}$. The results indicate that the timed *S* strategy is not spanned by the original *S* strategy. Thus, the dispersion of NS slope betas enhances the performance of the original *S* strategy.

[Insert Figure 3 and Table 8 around here]

*5.4.   Subsample and sector analyses*

We now examine the performance of the *S* strategy over various subsamples: before and after the introduction of the Commodity Futures Modernization Act (CFMA) in December 2000[17] and since the depths of the global financial crisis (GFC) dated March 2009. Figure 4, Panel A presents the Sharpe ratios of the *S* strategy for these different subsample periods, as well as those of the traditional commodity portfolios with positive mean excess returns (in Table 2). We note a tendency for the *S* strategy to perform strongly in the pre-CFMA sub-period and to show a weaker performance thereafter. This pattern is pervasive across all strategies except for Curve-M.

We compare the net Sharpe ratios and net alphas of the *S* portfolio pre- and post-event (namely, pre- and post-financialization and pre- and post-GFC). The pre-event net Sharpe ratios (1.74 for financialization and 1.47 for GFC) are statistically different from the post-event net Sharpe ratios (0.71 and 0.20, respectively) at the 1% level. Likewise, the annualized pre-event net alphas (2.16% and 1.90%, respectively) are statistically different from the annualized post-event net alphas (0.87% and 0.23%, respectively). These results, alongside those illustrated in Figure 4, indicate a significant deterioration of the net performance of the *S* strategy in latter periods.

────────────────

[17] The CFMA legislation led to the continued deregulation of OTC derivatives which increased their use in hedging and speculation in commodity markets. This effect caused a growth in trading volume and the financialization of commodity futures markets (Tang and Xiong, 2012).

The post-2001 sample includes the global financial crisis; therefore, we cannot attribute the observed decline in performance solely to the CFMA event. The decrease in performance may also be the result of crowding (Kang et al., 2021), market efficiency gains driven by algorithmic trading (Rösch et al., 2017) or the reduction of arbitrage limits in the latter period. It is conceivable that liquidity constraints and/or capacity constraints leading up to the introduction of the CFMA led to persistent mispricing of the futures curve that were subsequently arbitraged away as trading became cheaper and easier post-2001, or as liquidity rose and as frictions and funding costs became relatively lower.[18] This explanation is consistent with earlier evidence showing that the performance of the *S* strategy declines when the short-term rate is lower (see Table 7, Panel A).

[Insert Figure 4 around here]

As a further robustness check, we study the performance of the *S* strategy per sector (energy, grains, industrials, meats, metals, oilseeds, and softs). Figure 4, Panel B reports the Sharpe ratios of the sector-specific and overall portfolios, as well as *t*-statistics for the significance of the mean excess returns in parentheses. Except for meats, all sector-specific *S* strategies perform well with Sharpe ratios ranging from 0.36 (metals) to 1.24 (oilseeds) and corresponding *t*-statistics ranging from 2.11 to 6.44. The Sharpe ratio of the overall *S* portfolio is 1.41 and the average Sharpe ratio across sectors equals 0.60 with the difference highlighting the benefit of diversification.

*5.5. Variation in the modelling of the NS slope signal*

This section tests whether the performance of the *S* strategy is robust to three alternative designs of the NS parameters. First, we use extended commodity term structures with up to 6 or 12 contracts (in place of the 4 contracts originally considered), then we re-estimate Equation (1), and implement the $S$ strategy in relation to the front and most distant contracts available at the time of portfolio formation. The performance of the resulting strategies is summarized in Table 9, Panel A. Second, we extract the slope signal, $\beta_{S,t}$, from the following seasonally-adjusted NS model (Karstanje *et al.*, 2017)

$$F_t(M)$$

$$= \beta_{L,t} + \beta_{S,t}\left(\frac{1 - e^{-\lambda_t M}}{\lambda_t M}\right)$$

$$+ \beta_{C,t}\left(\frac{1 - e^{-\lambda_t M}}{\lambda_t M} - e^{-\lambda_t M}\right)$$

$$+ \beta_{SE,t}\cos(\omega M - \omega\theta) + \varepsilon_{t,M} \qquad (6)$$

where $\beta_{SE,t}$ is the seasonality beta of a given commodity at time $t$, $\omega = \frac{2\pi}{12}$, $\theta$ is the integer between 1 and 12 that maximizes the $R^2$ of the model at each point in time $t$, and the other parameters are as previously defined. Equation (6) is estimated using the term structure of each commodity up to the 12[th] contract with the $S$ strategy trading the front and most distant contracts available at the time of portfolio formation based on the estimated $\beta_{S,t}$ coefficients. The performance of the resulting $S$ strategy is reported in Table 9, Panel B with "NINE" ("ALL") indicating that the curves of nine (all) commodities are seasonality-adjusted.[19] Third, going back to the original setting of Equation (1) as applied to the four front contracts, we mitigate the noise in the slope signal by replacing it with its 3 or 5-day moving average (MA) and we present summary statistics of performance in Table 9, Panel C.

---

[18] Barroso and Detzel (2021) proxy arbitrage risk using idiosyncratic volatility, and proxy short-sell impediments using institutional ownership. Unlike the stock market, we refer to limits-to-arbitrage as the market's inability to absorb a large amount of capital in the short-term, i.e., a liquidity or capacity-constraint, rather than a structural impediment.

[Insert Table 9 around here]

---

[19] The nine commodity futures markets are corn, cotton, feeder cattle, gasoline, heating oil, live cattle, live hogs, soybeans, and wheat. These commodities are selected for their stronger seasonality characteristics. The $\beta_{S,t}$ coefficients of the remaining commodities are estimated from Equation (1).

Our conclusion regarding the strong performance of the *S* strategy is robust to these alternative specifications of the *S* signals. For example, the mean excess returns reported in Table 9 remain positive at the 1% level and, with an average at 1.65%, they are comparable to the mean excess return of the original *S* strategy (1.77% in Table 4). The CER (at 1.59% a year on average in Table 9) is also close to the CER of the original *S* strategy (at 1.74% in Table 4). Yet, we note that the longer term structure, a seasonality adjustment or smooth signals result in a marginal loss in risk-adjusted performance with, for example, Sharpe ratios that are slightly lower at 1 on average (in Table 9) versus 1.40 for the original *S* strategy (in Table 4).

### 5.6. *Overlay strategies*

We now examine whether the proposed *S* strategy can be used to hedge the risk of existing commodity risk factors and the potential benefits of using it as an overlay to traditional commodity allocations. We construct portfolios consisting of an equal monthly-rebalanced investment in *S* and either one of the seven traditional commodity portfolios presented in Table 2, Panel B. Results show that, net of reasonable transaction costs, treating *S* as an overlay improves the Sharpe ratios of traditional strategies (except for C-MOM) by an average of 22%.[20] Overall, the *S* strategy adds economic value both as a stand-alone portfolio and as a complement to traditional commodity portfolios.

### 5.7. *Alternative definitions of the slope signals*

Thus far, our modelling of the slope signal relies strictly on the NS model of Equation (1) or (6). We now determine the slope signal without relying on the NS model and use alternative slope measures as sorting signals for portfolio construction. Consistent with our NS-based *S* strategy, all three alternative strategies considered are based on daily rebalancing, full collateralization, as well as equal weighting in the construction of the long and short positions. The first two strategies also assume short-term continuation of recent slope movements, while the third definition anticipates no change in futures curves from one day to the next.[21] Details of the three alternative slope signals are as follows.

---

[20] In the interest of brevity, these results are not reported, but are available upon request.

The first alternative slope signal is the day $t$ change in the slope of the term structure of a given commodity, denoted $\Delta S_t$, with the slope measured as $S_t = F_t^1 - F_t^4$ and with $F_t^k$ denoting the futures price at time $t$ of a contract with $k^{th}$ location on the term structure. The strategy constructs a long slope-spread position (i.e., long the front contract and short the fourth contract) in the commodities with positive $\Delta S_t$ and short slope-spread positions in the remaining commodities (i.e., short the front contract and long the fourth contract).

The second alternative slope signal we consider is based on principal component analysis. Using the first four contracts on a given commodity over a five-day period ($t_1 = 1, \dots, t$ and $t = 5$), we first extract the second principal component, $PC_{2,t_1}$, from the covariance matrix of daily prices, where $PC_2$ is deemed to capture the slope of the futures curve. We then measure the change in $PC_2$ at time $t$ that we denote $\Delta PC_{2,t}$. A positive $\Delta PC_{2,t}$ indicates a more backwardated or less contangoed market and, vice versa, a negative $\Delta PC_{2,t}$ indicates a less backwardated or more contangoed market. Bearing in mind our assumption of short-term continuation of recent slope movements, we take long slope-spread positions in commodities with positive $\Delta PC_{2,t}$ and short slope-spread positions in commodities with negative $\Delta PC_{2,t}$.

The third alternative slope signal is defined as $RY_t = F_t^1/F_t^k - 1$ for $k = \{2, 3, 6, 12\}$ where $RY_t$ is the day $t$ roll yield of a given commodity. At each portfolio formation time, the strategy excludes the commodities with less than $k$ contracts in their term structure and takes long (short) positions in front contracts with positive (negative) roll yields.

---

[21] The assumption that the term structure of futures prices will not change from one portfolio formation date to the next (in our case, from one day to the next) follows the literature on carry strategies (see, for example, Erb and Harvey, 2006; Gorton and Rouwenhorst, 2006; Yang, 2013; Szymanowska et al., 2014).

Table 10 presents summary statistics for the excess returns of these alternative slope strategies (Panel A) as well as estimated coefficients from spanning regressions of the excess returns of the NS-based $S$ strategy onto the excess returns of the alternative slope strategies (Panel B). We note that all slope strategies are profitable at the 5% level or better, which confirms that the slope is important in the pricing of commodity futures contracts. We also observe that the NS-$S$ strategy presents positive alphas at the 1% level relative to the alternative slope strategies, which serves to confirm the overall robustness of our approach.

[Insert Table 10 around here]

## 6. Conclusions

This paper employs the Nelson-Siegel model to investigate the pricing of commodity futures contracts along their term structure. We first note that the model accurately captures the shape of commodity futures curves through the three parameters of level, slope and curvature. We then exploit the information embedded in these parameters within a set of novel out-of-sample investment strategies that assume continuation one day-ahead of recent parallel shifts ($L$ strategy), slope movements ($S$ strategy), or butterfly oscillations ($C$ strategy). The trades are implemented on a cross section of 21 commodities by constructing long and short positions in front contracts, slope spreads or butterfly spreads.

The proposed slope strategy generates superior excess returns that survive reasonable transaction costs and are uncorrelated to a variety of traditional commodity risk factors. Exploring the sources of the observed performance, we note that the profits of the slope strategy relate both to risk (since they compensate investors for the losses incurred during economic slowdowns) and to sentiment-based mispricing (i.e., due to the reluctance of traders to construct short positions in high sentiment periods). Further tests indicate that the performance of the slope strategy remains robust and persistent under alternative specifications of the Nelson-Siegel model and can be enhanced through timing. Yet, performance is weaker in the recent past, which we attribute to recent technological advances, strategy crowding and/or an accommodative monetary policy environment.

Our article is the first to employ the NS model to design novel investment strategies in commodity futures markets. We hope this research motivates further investigation of how fixed income term structure models can be used to enhance our understanding of the pricing and management of futures contracts along their entire curves and across asset classes.

*Acknowledgements*

We thank Geert Bekaert (editor), the associate editor, an anonymous referee, as well as Andrew Kaleel, Mathew Kaleel, Bin Li, Mingyi Li, Ming Xu and seminar participants at Griffith University for insightful comments and suggestions.

# References

Asness, C. S., Moskowitz, T. J., Pedersen, L. H., 2013. Value and momentum everywhere. The Journal of Finance, 68 (3), 929-985.

Baker, M., Wurgler, J., 2006. Investor sentiment and the cross-section of stock returns. The Journal of Finance, 61 (4), 1645–1680.

Bakshi, G., Gao, X., Rossi, A. G., 2019. Understanding the sources of risk underlying the cross section of commodity returns. Management Science, 65 (2), 619-641.

Barroso, P., Detzel, A., 2021. Do limits to arbitrage explain the benefits of volatility-managed portfolios?. Journal of Financial Economics, 140 (3), 744-767.

Basu, D., Miffre, J., 2013. Capturing the risk premium of commodity futures: The role of hedging pressure. Journal of Banking and Finance, 37 (7), 2652-2664.

Bekaert, G., Engstrom, E. C., Xu, N. R., 2022. The time variation in risk appetite and uncertainty. Management Science, 68 (6), 3975-4004.

Birru, J., 2018. Day of the week and the cross-section of returns. Journal of Financial Economics, 130 (1), 182-214.

Boons, M., Prado, M. P., 2019. Basis-Momentum. The Journal of Finance, 74 (1), 239-279.

Carhart, M. M., 1997. On persistence in mutual fund performance. The Journal of Finance, 52 (1), 57–82.

Casassus, J., Collin-Dufresne, P., 2005. Stochastic convenience yield implied from commodity futures and interest rates. The Journal of Finance, 60 (5), 2283-2331.

Deaton, A., Laroque, G., 1992. On the behaviour of commodity prices. Review of Economic Studies, 59 (1), 1-23.

Diebold, F. X., Li, C., 2006. Forecasting the term structure of government bond yields. Journal of Econometrics, 130 (2), 337-364.

Dunis, C., Laws, J., Evans, B., 2006. Trading futures spreads: An application of correlation and threshold filters. Applied Financial Economics, 16 (12), 903-914.

Erb, C. B., Harvey, C. R., 2006. The strategic and tactical value of commodity futures. Financial Analysts Journal, 62 (2), 69-97.

Fama, E.F., French, K.R., 2015. A five-factor asset pricing model. Journal of Financial Economics, 116 (1), 1–22.

Fernandez-Perez, A., Frijns, B., Fuertes, A.-M., Miffre, J., 2018. The skewness of commodity futures returns. Journal of Banking and Finance, 86, 143-158.

Gao, L., Han, Y., Li, S. Z., Zhou, G., 2018. Market intraday momentum. Journal of Financial Economics, 129 (2), 394-414.

Gibson, R., Schwartz, E. S., 1990. Stochastic convenience yield and the pricing of oil contingent claims. The Journal of Finance, 45 (3), 959-976.

Gorton, G., Rouwenhorst, K. G., 2006. Facts and fantasies about commodity futures. Financial Analysts Journal, 62 (2), 47-68.

Gorton, G. B., Hayashi, F., Rouwenhorst, K. G., 2013. The fundamentals of commodity futures returns. Review of Finance, 17 (1), 35-105.

GrØnborg, N. S., Lunde, A., 2016. Analyzing oil futures with a dynamic Nelson‑Siegel model. Journal of Futures Markets, 36 (2), 153-173.

Gu, M., Kang, W., Lou, D., Tang, K., 2019. Relative basis and expected returns in commodity futures markets. SSRN working paper.

Heath, D., 2019. Macroeconomic factors in oil futures markets. Management Science, 65 (9), 4407-4421.

Heidorn, T., Mokinski, F., Rühl, C., Schmaltz, C., 2015. The impact of fundamental and financial traders on the term structure of oil. Energy Economics, 48, 276-287.

Ilmanen, A., Israel, R., Lee, R., Moskowitz, T., Thapar, A., 2021. How do factor premia vary over time? A century of evidence. Journal of Investment Management, 19 (4), 15-57.

Jurado, K., Ludvigson, S. C., Ng, S., 2015. Measuring uncertainty. American Economic Review, 105 (3), 1177-1216.

Kang, W., Rouwenhorst, K. G., Tang, K., 2020. A tale of two premiums: The role of hedgers and speculators in commodity futures markets. The Journal of Finance, 75 (1), 377- 417.

Kang, W., Rouwenhorst, K. G., Tang, K., 2021. Crowding and factor returns. SSRN working paper.

Karstanje, D., Van Der Wel, M., van Dijk, D. J., 2017. Common factors in commodity futures curves. SSRN working paper.

Koijen, R. S. J., Moskowitz, T. J., Pedersen, L. H., Vrugt, E. B., 2018. Carry. Journal of Financial Economics, 127 (2), 197-225.

Locke, P. R., Venkatesh, P., 1997. Futures market transaction costs. Journal of Futures Markets, 17 (2), 229-245.

Melamed, L., 1981. The futures market: Liquidity and the technique of spreading. Journal of Futures Markets, 1 (3), 405.

Miffre, J., Rallis, G., 2007. Momentum strategies in commodity futures markets. Journal of Banking and Finance, 31 (6), 1863-1886.

Moskowitz, T. J., Ooi, Y. H., Pedersen, L. H., 2012. Time series momentum. Journal of Financial Economics, 104 (2), 228-250.

Nelson, C. R., Siegel, A. F., 1987. Parsimonious modeling of yield curves. The Journal of Business, 60 (4), 473-489.

Newey, W. K., West, K. D., 1987. A simple, positive semi-definite, heteroskedasticity and autocorrelation consistent covariance matrix. Econometrica, 55 (3), 703-708.

O'Neill, M., Schmidt, C., Warren, G., 2018. Capacity Analysis for Equity Funds. The Journal of Portfolio Management, 44 (5), 36-49.

Paschke, R., Prokopczuk, M., Wese Simen, C., 2020. Curve momentum. Journal of Banking and Finance, 113, 105718.

Pastor, L., Stambaugh, R. F., 2003. Liquidity risk and expected stock returns. Journal of Political Economy, 111 (3), 642-685.

Rösch, D. M., Subrahmanyam, A., Van Dijk, M. A., 2017. The dynamics of market efficiency. Review of Financial Studies, 30 (4), 1151-1187.

Schwartz, E. S., 1997. The stochastic behavior of commodity prices: Implications for valuation and hedging. The Journal of Finance, 52 (3), 923-973.

Stambaugh, R. F., Yu, J., Yuan, Y., 2012. The short of it: Investor sentiment and anomalies. Journal of Financial Economics, 104 (2), 288-302.

Szakmary, A. C., Shen, Q., Sharma, S. C., 2010. Trend-following trading strategies in commodity futures: A re-examination. Journal of Banking and Finance, 34 (2), 409-426.

Szymanowska, M., de Roon, F. A., Nijman, T. E., van den Goorbergh, R. W. J., 2014. An anatomy of commodity futures risk premia. The Journal of Finance, 69 (1), 453-482.

Tang, K., Xiong, W., 2012. Index investment and the financialization of commodities. Financial Analysts Journal, 68 (5), 54-74.

Yang, F., 2013. Investment shocks and the commodity basis spread. Journal of Financial Economics, 110 (1), 164-184.

Zaremba, A., Umutlu, M., Maydybura, A., 2020. Where have the profits gone? Market efficiency and the disappearing equity anomalies in country and industry returns. Journal of Banking and Finance, 121, 105966.

# Appendix A. Factor Construction

| Factor | Commodity-specific signals | Definition at the time of portfolio formation $t$ | References |
|---|---|---|---|
| Momentum (MOM) | $MOM_t = \prod_{s=0}^{11}(1 + r_{t-s}) - 1$ | $r_t$ denotes the time $t$ monthly excess return of the front contract. | Erb and Harvey (2006); Miffre and Rallis (2007); Bakshi et al. (2019) |
| Carry (Carry) | $RY_t = F_t^1 / F_t^2 - 1$ | $F_t^1$ and $F_t^2$ denote the prices of the nearest and $2^{nd}$ nearest contract at time $t$, respectively. | Erb and Harvey (2006); Gorton and Rouwenhorst (2006); Szymanowska et al. (2014); Yang (2013); Bakshi et al. (2019) |
| Hedging Pressure (HP) | $HP_t = \frac{1}{52}\sum_{w=0}^{51}\frac{S_{t-w} - L_{t-w}}{S_{t-w} + L_{t-w}}$ | $S_t$ and $L_t$ correspond to the week $t$ short and long positions of a given commodity as held by commercial traders in the CFTC report, respectively. | Basu and Miffre (2013); Kang et al. (2020) |
| Skewness (SKEW) | $SKEW_t = \left\{\left[\frac{1}{D_1}\sum_{d=0}^{D_1-1}(r_{t-d} - \mu_t)^3\right] \middle/ \sigma_t^3\right\}$ | $r_d$ denotes the daily excess return of the front contract at time $d$, $\mu_t$ and $\sigma_t$ denote mean and standard deviation of daily excess returns as measured at time $t$ using daily data over the past year and $D_1$ is the number of days in the past one year. | Fernandez-Perez et al. (2018) |
| Basis-momentum (BMOM) | $BM = \prod_{s=0}^{11}(1 + r_{t-s}^1) - \prod_{s=0}^{11}(1 + r_{t-s}^2)$ | $r_t^1$ ($r_t^2$) represents the time $t$ monthly excess return of the front (second-nearest) contract. | Boons and Prado (2019) |
| Relative basis (RB) | $RB_t = \frac{ln(F_t^1/F_t^2)}{T_t^2 - T_t^1} - \frac{ln(F_t^2/F_t^3)}{T_t^3 - T_t^2}$ | $F_t^m$ denotes the time $t$ price of the $m$th nearest contract, $T_t^m$ represents the time to maturity of the $m$th nearest contract expressed in number of days at time $t$. | Gu et al. (2019) |
| Liquidity (LIQ) | $LIQ_t = \frac{1}{D_2}\sum_{d=0}^{D_2-1}\frac{\$Vol_{t-d}}{|r_{t-d}|}$ | $\$Vol_t$ and $r_t$ denote the day $t$ dollar volume and excess return of the front contract, respectively. $D_2$ is the number of days in the past 2 months. | Szymanowska et al. (2014) |
| Curve-momentum (Curve-M) | $MOM_t = \prod_{s=0}^{11}(1 + r_{t-s}^m) - 1$ | $r_t^m$ represents the excess return of the $m$th contract at time $t$. | Paschke et al. (2020) |

**Table 1**. **Summary Statistics of Commodity Excess Returns**

The table reports summary statistics for the excess returns of individual commodity futures categorized into seven sectors (Panel A) and cross section (Panel B). Mean and SD are the annualized mean and annualized standard deviation of excess returns, respectively. M is the location of the contract on the term structure. The long slope spread (Spread) is defined as a long position in the front and a short position in the 4th contract, while the long butterfly spread (Butterfly) is constructed with two long positions in the 2nd contract, one short position in the front contract and one short position in the 4th contract. The last column reports the average daily ratio of the cumulative open interest of the first 4 contracts (COI) to the total open interest of the entire curve (TOI). Numbers in bold format indicate statistical significance at the 10% level or above. *t*-statistics in parentheses test the significance of the mean. The last two rows present the percentages of positive (negative) mean returns at the 10% level. The sample covers the period from January 1992 to June 2019.

| Sector | Commodity | M=1 | | M=2 | | M=3 | | M=4 | | Spread | | Butterfly | | COI/TOI |
|---|---|---|---|---|---|---|---|---|---|---|---|---|---|---|
| | | Mean | SD | Mean | SD | Mean | SD | Mean | SD | Mean | SD | Mean | SD | |
| **Panel A: Individual commodity summary statistics** | | | | | | | | | | | | | | |
| Energy | Crude oil | 0.0145 (1.15) | 0.3292 | 0.0341 (1.46) | 0.3114 | 0.0446 (1.64) | 0.2971 | **0.0511** (1.78) | 0.2859 | -0.0243 (-1.45) | 0.0744 | 0.0028 (0.62) | 0.0290 | 44.03% |
| | Gasoline | **0.0724** (2.08) | 0.3208 | **0.0684** (2.03) | 0.2968 | **0.0696** (2.10) | 0.2818 | **0.0726** (2.20) | 0.2720 | 0.0087 (0.70) | 0.0908 | -0.0075 (-0.90) | 0.0396 | 58.03% |
| | Heating oil | 0.0264 (1.28) | 0.3056 | 0.0303 (1.33) | 0.2890 | 0.0358 (1.43) | 0.2763 | 0.0430 (1.58) | 0.2666 | -0.0100 (-0.55) | 0.0698 | **-0.0102** (-1.82) | 0.0321 | 52.67% |
| Grains | Corn | -0.0751 (-0.91) | 0.2483 | -0.0616 (-0.69) | 0.2405 | -0.0415 (-0.33) | 0.2307 | -0.0327 (-0.20) | 0.2203 | **-0.0393** (-2.29) | 0.0819 | -0.0153 (-0.94) | 0.0715 | 87.02% |
| | Oats | -0.0096 (0.62) | 0.3067 | -0.0247 (0.23) | 0.2720 | -0.0283 (0.06) | 0.2496 | -0.0264 (0.03) | 0.2348 | 0.0187 (1.01) | 0.1871 | -0.0284 (-0.58) | 0.1665 | 95.79% |
| | Rough rice | **-0.1083** (-1.64) | 0.2378 | -0.0777 (-1.16) | 0.2212 | -0.0579 (-0.85) | 0.2066 | -0.0358 (-0.42) | 0.1947 | **-0.0762** (-3.06) | 0.1194 | -0.0197 (-0.61) | 0.1233 | 96.44% |
| | Wheat | -0.1057 (-1.22) | 0.2834 | -0.0808 (-0.89) | 0.2665 | -0.0506 (-0.39) | 0.2521 | -0.0385 (-0.24) | 0.2370 | **-0.0610** (-2.72) | 0.1063 | -0.0183 (-0.99) | 0.0862 | 93.49% |
| Industrials | Cotton | -0.0385 (-0.09) | 0.2597 | -0.0376 (-0.21) | 0.2360 | -0.0189 (0.12) | 0.2193 | -0.0204 (0.01) | 0.2048 | -0.0148 (-0.22) | 0.1334 | -0.0218 (-0.65) | 0.1255 | 89.31% |
| | Lumber | -0.0855 (-0.70) | 0.2889 | -0.0555 (-0.40) | 0.2589 | -0.0389 (-0.27) | 0.2294 | -0.0148 (0.17) | 0.2101 | -0.0684 (-1.41) | 0.1947 | -0.0236 (-0.21) | 0.1916 | 96.41% |
| Meats | Feeder cattle | 0.0160 (0.95) | 0.1423 | 0.0102 (0.74) | 0.1396 | 0.0272 (1.43) | 0.1322 | **0.0350** (1.80) | 0.1247 | -0.0181 (-1.58) | 0.0532 | **-0.0283** (-2.81) | 0.0575 | 86.04% |
| | Live cattle | 0.0060 (0.61) | 0.1461 | 0.0199 (1.18) | 0.1304 | 0.0142 (1.01) | 0.1135 | 0.0184 (1.29) | 0.1038 | -0.0106 (-0.44) | 0.0838 | 0.0133 (1.22) | 0.0779 | 91.56% |
| | Live hogs | -0.0495 (-0.39) | 0.2441 | -0.0241 (0.01) | 0.2219 | 0.0113 (0.81) | 0.1941 | 0.0202 (1.08) | 0.1749 | -0.0698 (-1.50) | 0.1720 | -0.0297 (-0.52) | 0.1719 | 85.26% |
| Metals | Copper | 0.0437 (1.64) | 0.2588 | 0.0424 (1.62) | 0.2569 | **0.0467** (1.71) | 0.2534 | **0.0511** (1.81) | 0.2500 | -0.0060 (-0.89) | 0.0341 | **-0.0093** (-2.09) | 0.0271 | 73.48% |
| | Gold | 0.0232 (1.19) | 0.1600 | 0.0240 (1.21) | 0.1598 | 0.0242 (1.22) | 0.1598 | 0.0245 (1.23) | 0.1596 | -0.0010 (-0.89) | 0.0062 | 0.0006 (0.73) | 0.0104 | 82.13% |
| | Silver | 0.0174 (1.10) | 0.2845 | 0.0177 (1.11) | 0.2835 | 0.0219 (1.19) | 0.2824 | 0.0215 (1.18) | 0.2813 | **-0.0053** (-1.97) | 0.0145 | -0.0020 (-1.21) | 0.0134 | 86.42% |
| Oilseeds | Soybean meal | **0.1015** (2.68) | 0.2468 | **0.0708** (2.15) | 0.2393 | **0.0652** (2.05) | 0.2348 | **0.0618** (2.00) | 0.2311 | **0.0383** (2.05) | 0.1005 | **-0.0243** (-1.75) | 0.0739 | 78.45% |
| | Soybean oil | -0.0331 (-0.21) | 0.2205 | -0.0322 (-0.20) | 0.2174 | -0.0191 (0.09) | 0.2137 | -0.0111 (0.21) | 0.2104 | **-0.0213** (-2.51) | 0.0534 | **-0.0205** (-3.47) | 0.0478 | 79.66% |
| | Soybeans | 0.0405 (1.53) | 0.2218 | 0.0287 (1.25) | 0.2177 | 0.0342 (1.39) | 0.2149 | 0.0365 (1.46) | 0.2094 | 0.0040 (0.48) | 0.0725 | **-0.0214** (-1.85) | 0.0612 | 80.99% |
| Softs | Cocoa | -0.0116 (0.56) | 0.2932 | -0.0292 (0.18) | 0.2779 | -0.0230 (0.25) | 0.2677 | -0.0210 (0.26) | 0.2606 | 0.0149 (1.34) | 0.0709 | **-0.0282** (-3.06) | 0.0608 | 80.35% |
| | Coffee | -0.0572 (0.16) | 0.3707 | -0.0639 (-0.06) | 0.3474 | -0.0721 (-0.31) | 0.3258 | -0.0606 (-0.19) | 0.3151 | 0.0150 (1.17) | 0.1203 | -0.0126 (-0.53) | 0.1005 | 92.20% |
| | Orange juice | -0.0754 (-0.42) | 0.3164 | -0.0881 (-0.85) | 0.2826 | -0.0848 (-0.94) | 0.2650 | -0.0820 (-0.98) | 0.2545 | 0.0147 (1.12) | 0.1347 | -0.0285 (-1.29) | 0.1258 | 95.51% |
| **Panel B: Cross-section summary statistics** | | | | | | | | | | | | | | |
| Cross-section | | -0.0137 | 0.2612 | -0.0109 | 0.2460 | -0.0019 | 0.2334 | 0.0044 | 0.2239 | -0.0149 | 0.0940 | -0.0159 | 0.0806 | 82.16% |
| % of positive mean return | | 10% | | 10% | | 14% | | 24% | | 5% | | 0% | | |
| % of negative mean return | | 0% | | 0% | | 0% | | 0% | | 24% | | 33% | | |

**Table 2. Summary Statistics of the Risk Factors**

The table reports performance statistics for benchmark portfolios. Panel A focuses on naive commodity benchmarks that are long-only, daily-rebalanced portfolios of level (LAVG), slope (SAVG) or curvature (CAVG) positions. Panel B presents the traditional commodity risk premia (AVG is a long-only monthly-rebalanced portfolio of all commodities, MOM, CARRY, HP, SKEW, BMOM, RB, LIQ, and Curve-M denote long-short portfolios sorted on the characteristics presented in Appendix A). Mean, Volatility, Downside volatility, and CER (certainty equivalent return) are annualized. Newey and West (1987) adjusted *t*-statistics are reported in parentheses. The sample covers the period from January 1992 to June 2019.

| | Panel A: Naïve commodity benchmarks | | | Panel B: Traditional commodity risk factors | | | | | | | | |
| --- | --- | --- | --- | --- | --- | --- | --- | --- | --- | --- | --- | --- |
| | LAVG | SAVG | CAVG | AVG | MOM | CARRY | HP | SKEW | BMOM | RB | LIQ | Curve-M |
| Mean | 0.0151 | -0.0046 | -0.0027 | 0.0158 | 0.0264 | 0.0201 | 0.0355 | 0.0283 | 0.0253 | 0.0274 | 0.0008 | 0.0055 |
| | (0.97) | (-1.71) | (-2.99) | (0.89) | (2.11) | (1.83) | (2.92) | (2.47) | (2.39) | (2.44) | (0.23) | (5.65) |
| Volatility | 0.1174 | 0.0136 | 0.0056 | 0.1201 | 0.0700 | 0.0648 | 0.0638 | 0.0655 | 0.0553 | 0.0621 | 0.0640 | 0.0051 |
| Downside volatility | 0.0840 | 0.0085 | 0.0037 | 0.0862 | 0.0452 | 0.0407 | 0.0385 | 0.0395 | 0.0300 | 0.0371 | 0.0365 | 0.0029 |
| Sharpe ratio | 0.1285 | -0.3365 | -0.4881 | 0.1316 | 0.3774 | 0.3097 | 0.5559 | 0.4320 | 0.4571 | 0.4413 | 0.0129 | 1.0760 |
| Sortino ratio | 0.2643 | -0.5262 | -0.7346 | 0.2706 | 0.6474 | 0.5501 | 0.9905 | 0.7822 | 0.9030 | 0.7996 | 0.0785 | 1.8621 |
| Omega ratio | 1.0299 | 0.9450 | 0.9202 | 1.1251 | 1.3378 | 1.2795 | 1.5585 | 1.4046 | 1.4950 | 1.4164 | 1.0135 | 2.2556 |
| Skewness | -0.2804 | 0.2577 | -0.0458 | -0.4644 | -0.2523 | -0.0923 | -0.1148 | 0.0653 | 0.3227 | 0.1492 | 0.1430 | 0.0843 |
| Excess kurtosis | 5.2614 | 2.4187 | 2.6389 | 3.9242 | 0.8004 | 0.7519 | -0.0691 | 0.3920 | 0.4682 | 0.8402 | 0.0206 | 0.6692 |
| 99% VaR (Cornish-Fisher) | 0.0245 | 0.0027 | 0.0010 | 0.1000 | 0.0482 | 0.0469 | 0.0435 | 0.0496 | 0.0460 | 0.0503 | 0.0460 | 0.0042 |
| % of positive months | 0.5107 | 0.4862 | 0.4824 | 0.5167 | 0.5426 | 0.5319 | 0.5912 | 0.5678 | 0.5521 | 0.5167 | 0.4909 | 0.6530 |
| Maximum drawdown | -0.4736 | -0.1728 | -0.0888 | -0.4352 | -0.2872 | -0.2050 | -0.1529 | -0.1607 | -0.2195 | -0.1536 | -0.3378 | -0.0071 |
| CER | -0.0126 | -0.0049 | -0.0028 | -0.0149 | 0.0165 | 0.0116 | 0.0272 | 0.0197 | 0.0192 | 0.0197 | -0.0073 | 0.0054 |

**Table 3**. **Estimates from the Nelson-Siegel Model**

The table presents summary statistics for the Nelson and Siegel (1987) parameters from Equation (1). Mean and SD are the average and standard deviation of the estimated coefficients, respectively. AC(1) are $p$-values for the hypothesis of no first-order correlation in the change in beta. The right-hand side of the table reports the average $R^2$ (expressed as a percentage) of various specifications of the NS model where L, S and C refer to the level, slope, and curvature components, respectively. The sample covers the period from January 1992 to June 2019.

| Commodity | | Level beta | | | Slope beta | | | Curvature beta | | | Decay factor | | Average $R^2$ (%) | | |
|---|---|---|---|---|---|---|---|---|---|---|---|---|---|---|---|
| | | Mean | SD | AC(1) | Mean | SD | AC(1) | Mean | SD | AC(1) | Mean | SD | L + S + C | L + S | L + C |
| Energy | Crude oil | 49.63 | 30.72 | 0.03 | -0.90 | 7.43 | 0.00 | 2.04 | 8.60 | 0.00 | 0.45 | 0.03 | 99.70 | 93.07 | 24.86 |
| | Gasoline | 1.38 | 1.05 | 0.98 | 0.07 | 0.50 | 0.23 | 0.06 | 2.03 | 0.83 | 0.45 | 0.03 | 96.20 | 76.23 | 31.24 |
| | Heating oil | 1.48 | 0.96 | 0.01 | -0.04 | 0.22 | 0.03 | 0.01 | 0.53 | 0.02 | 0.45 | 0.03 | 98.40 | 83.92 | 26.02 |
| Grains | Corn | 3.62 | 1.36 | 0.24 | -0.27 | 0.87 | 0.38 | 0.00 | 2.35 | 0.72 | 0.34 | 0.07 | 97.00 | 84.07 | 42.22 |
| | Oats | 2.25 | 1.00 | 0.01 | -0.12 | 0.62 | 0.44 | -0.07 | 1.40 | 0.31 | 0.33 | 0.05 | 96.30 | 81.13 | 43.67 |
| | Rough rice | 10.78 | 4.06 | 0.00 | -0.87 | 2.03 | 0.13 | -0.18 | 6.07 | 0.00 | 0.37 | 0.05 | 94.50 | 81.33 | 38.91 |
| | Wheat | 4.82 | 2.30 | 0.00 | -0.38 | 1.46 | 0.00 | 0.19 | 3.99 | 0.00 | 0.33 | 0.05 | 96.40 | 83.31 | 38.98 |
| Industrials | Cotton | 0.68 | 0.18 | 0.15 | -0.01 | 0.26 | 0.00 | 0.04 | 0.49 | 0.31 | 0.33 | 0.05 | 95.50 | 76.43 | 41.35 |
| | Lumber | 321.04 | 57.57 | 0.00 | -21.85 | 70.17 | 0.00 | -5.20 | 144.49 | 0.00 | 0.35 | 0.05 | 92.50 | 76.00 | 43.70 |
| Meats | Feeder cattle | 1.10 | 0.39 | 0.00 | 0.00 | 0.12 | 0.63 | -0.01 | 0.35 | 0.00 | 0.47 | 0.09 | 94.00 | 74.09 | 40.12 |
| | Live cattle | 0.92 | 0.34 | 0.34 | 0.00 | 0.19 | 0.20 | 0.03 | 0.67 | 0.30 | 0.35 | 0.05 | 90.40 | 61.82 | 41.59 |
| | Live hogs | 0.73 | 0.38 | 0.08 | -0.11 | 0.32 | 0.88 | -0.04 | 0.90 | 0.07 | 0.48 | 0.12 | 92.30 | 68.20 | 42.15 |
| Metals | Copper | 2.00 | 1.15 | 0.00 | 0.01 | 0.09 | 0.00 | 0.01 | 0.14 | 0.00 | 0.60 | 0.06 | 97.10 | 87.53 | 31.01 |
| | Gold | 776.43 | 470.59 | 0.16 | -21.68 | 14.87 | 0.00 | -23.31 | 16.89 | 0.00 | 0.36 | 0.04 | 99.90 | 96.38 | 33.03 |
| | Silver | 12.36 | 8.61 | 0.46 | -0.36 | 0.25 | 0.00 | -0.32 | 0.32 | 0.00 | 0.36 | 0.05 | 99.20 | 95.91 | 35.76 |
| Oilseeds | Soybean meal | 233.35 | 87.15 | 0.02 | 22.15 | 64.87 | 0.00 | 33.13 | 153.22 | 0.04 | 0.50 | 0.10 | 96.90 | 81.52 | 39.01 |
| | Soybean oil | 0.31 | 0.12 | 0.00 | -0.01 | 0.03 | 0.00 | 0.00 | 0.06 | 0.00 | 0.46 | 0.09 | 98.70 | 91.49 | 36.19 |
| | Soybeans | 7.91 | 2.92 | 0.02 | 0.36 | 1.75 | 0.09 | 0.78 | 4.43 | 0.26 | 0.44 | 0.09 | 96.10 | 79.64 | 38.37 |
| Softs | Cocoa | 2001.68 | 662.05 | 0.00 | -114.84 | 147.50 | 0.01 | -99.96 | 217.46 | 0.00 | 0.31 | 0.05 | 98.40 | 86.50 | 40.26 |
| | Coffee | 1.34 | 0.45 | 0.00 | -0.13 | 0.25 | 0.19 | -0.14 | 0.24 | 0.00 | 0.31 | 0.05 | 99.40 | 93.71 | 39.42 |
| | Orange juice | 1.26 | 0.27 | 0.00 | -0.08 | 0.18 | 0.00 | -0.12 | 0.28 | 0.00 | 0.36 | 0.04 | 97.60 | 84.55 | 38.98 |

**Table 4. Performance of Nelson-Siegel Portfolios (Summary Statistics)**

The table presents performance statistics for the long, short and long-short portfolios based on the Level (L), Slope (S), and Curvature (C) strategies. Mean, Volatility, Downside volatility, and CER (certainty equivalent return) are expressed as annualized terms. The mean is geometric. Newey and West (1987) adjusted *t*-statistics are reported in parentheses. The sample covers the period from January 1992 to June 2019.

| | **L** | | | **S** | | | **C** | | |
|---|---|---|---|---|---|---|---|---|---|
| | Long | Short | Long-Short | Long | Short | Long-Short | Long | Short | Long-Short |
| Mean | 0.0222 | 0.0214 | -0.0019 | 0.0160 | -0.0196 | 0.0177 | 0.0059 | -0.0078 | 0.0068 |
| | (1.26) | (1.17) | (0.01) | (4.13) | (-5.57) | (7.23) | (3.18) | (-5.24) | (5.34) |
| Volatility | 0.1297 | 0.1339 | 0.0630 | 0.0204 | 0.0170 | 0.0125 | 0.0088 | 0.0070 | 0.0055 |
| Downside volatility | 0.0881 | 0.0939 | 0.0392 | 0.0134 | 0.0111 | 0.0082 | 0.0064 | 0.0048 | 0.0037 |
| Sharpe ratio | 0.1716 | 0.1599 | -0.0300 | 0.7859 | -1.1493 | 1.4135 | 0.6642 | -1.1009 | 1.2263 |
| Sortino ratio | 0.3534 | 0.3288 | 0.0024 | 1.2237 | -1.7335 | 2.1998 | 0.9286 | -1.6064 | 1.8475 |
| Omega ratio | 1.0383 | 1.0374 | 1.0009 | 1.1478 | 0.8229 | 1.2689 | 1.1238 | 0.8280 | 1.2272 |
| Skewness | -0.0963 | -0.1189 | 0.1898 | 0.2675 | 0.2095 | 0.0039 | -0.2801 | -0.0091 | -0.1612 |
| Excess kurtosis | 3.3382 | 3.9847 | 1.9456 | 4.8917 | 2.8921 | 2.9167 | 7.8971 | 2.0969 | 3.2770 |
| 99% VaR (Cornish-Fisher) | 0.0262 | 0.0284 | 0.0118 | 0.0049 | 0.0034 | 0.0025 | 0.0022 | 0.0012 | 0.0011 |
| % of positive months | 0.5163 | 0.5100 | 0.4912 | 0.5905 | 0.5190 | 0.5986 | 0.5291 | 0.4671 | 0.5344 |
| Maximum drawdown | -0.4532 | -0.4759 | -0.3803 | -0.0707 | -0.4288 | -0.0271 | -0.0527 | -0.1599 | -0.0114 |
| CER | -0.0115 | -0.0146 | -0.0098 | 0.0152 | -0.0202 | 0.0174 | 0.0057 | -0.0079 | 0.0067 |

**Table 5. Turnover and Transaction Cost Analysis**

The table presents the turnover, annualized mean excess return (Mean) and Sharpe ratio (SR) of various strategies. The portfolios considered are the *S* and *C* portfolios, and seven long-short characteristic-sorted portfolios based on momentum (MOM), carry (CARRY), hedging pressure (HP), skewness (SKEW), basis-momentum (BMOM), relative basis (RB), and curve-momentum (Curve-M). TC1, TC2 and TC3 denote the three transaction cost scenarios as detailed in Section 4.3. Newey and West (1987) adjusted *t*-statistics are reported in parentheses. The sample covers the period from January 1992 to June 2019.

| | Turnover | Gross return | | Net return | | | | | |
| | | | | TC1 | | TC2 | | TC3 | |
| | | Mean | SR | Mean | SR | Mean | SR | Mean | SR |
| **Panel A: NS strategies** | | | | | | | | | |
| S | 1.17 | 0.0177 (7.23) | 1.41 | 0.0131 (5.50) | 1.04 | 0.0055 (2.31) | 0.44 | 0.0055 (2.34) | 0.44 |
| C | 1.20 | 0.0068 (5.34) | 1.23 | 0.0024 (2.17) | 0.43 | -0.0054 (-4.83) | -0.97 | -0.0055 (-4.98) | -0.98 |
| **Panel B: Traditional strategies** | | | | | | | | | |
| MOM | 1.34 | 0.0264 (2.11) | 0.38 | 0.0260 (2.08) | 0.37 | 0.0251 (2.01) | 0.36 | 0.0252 (2.03) | 0.36 |
| CARRY | 1.42 | 0.0201 (1.83) | 0.31 | 0.0196 (1.80) | 0.30 | 0.0186 (1.72) | 0.29 | 0.0188 (1.73) | 0.29 |
| HP | 1.25 | 0.0355 (2.92) | 0.56 | 0.0350 (2.89) | 0.55 | 0.0342 (2.83) | 0.54 | 0.0343 (2.84) | 0.54 |
| SKEW | 1.33 | 0.0283 (2.47) | 0.43 | 0.0278 (2.43) | 0.42 | 0.0270 (2.36) | 0.41 | 0.0271 (2.37) | 0.41 |
| BMOM | 1.32 | 0.0253 (2.39) | 0.46 | 0.0248 (2.35) | 0.45 | 0.0239 (2.27) | 0.43 | 0.0241 (2.28) | 0.44 |
| RB | 1.56 | 0.0274 (2.44) | 0.44 | 0.0268 (2.40) | 0.43 | 0.0258 (2.31) | 0.42 | 0.0259 (2.32) | 0.42 |
| Curve-M | 1.31 | 0.0055 (5.65) | 1.08 | 0.0050 (5.18) | 0.99 | 0.0042 (4.29) | 0.82 | 0.0043 (4.41) | 0.84 |

**Table 6. Risk Adjustment and Abnormal Performance of the *S* and *C* Strategies**

The table presents estimates and goodness-of-fit statistics (Adj-$R^2$) from regressions of the net excess returns of the *S* and *C* strategies on naive commodity benchmarks (Panel A), traditional commodity benchmarks (Panel B), and all-asset benchmarks (Panel C). LAVG, SAVG and CAVG are naive long-only, daily-rebalanced portfolios of the level, slope and curvature positions, respectively. AVG is a long-only, monthly-rebalanced portfolio of all commodities, MOM, CARRY, HP, SKEW, BMOM, RB, LIQ and Curve-M denote long-short commodity portfolios sorted on characteristics presented in Appendix A. Alpha is expressed in annualized terms. All variables are monthly series (or converted to monthly series where applicable). Newey and West (1987) adjusted *t*-statistics are reported in parentheses. The sample covers the period from January 1992 to June 2019.

| | **S** | | | | **C** | | | |
|---|---|---|---|---|---|---|---|---|
| **Panel A: Naive commodity benchmarks** | | | | | | | | |
| Alpha | 0.0139 | (4.56) | | | 0.0028 | (2.18) | | |
| LAVG | 0.0018 | (0.28) | | | 0.0002 | (0.05) | | |
| SAVG | -0.0008 | (-0.01) | | | 0.0029 | (0.11) | | |
| CAVG | 0.2099 | (1.26) | | | 0.0994 | (1.29) | | |
| Adj-$R^2$ | -0.35% | | | | -0.37% | | | |
| | | | | | | | | |
| **Panel B: Traditional commodity benchmarks** | | | | | | | | |
| Alpha | 0.0125 | (4.22) | 0.0131 | (4.13) | 0.0022 | (1.71) | 0.0019 | (1.49) |
| AVG | 0.0023 | (0.41) | 0.0052 | (0.87) | 0.0010 | (0.36) | 0.0007 | (0.27) |
| MOM | 0.0163 | (1.32) | 0.0177 | (1.21) | 0.0052 | (0.87) | 0.0038 | (0.54) |
| CARRY | 0.0074 | (0.52) | 0.0207 | (1.42) | -0.0050 | (-0.74) | -0.0078 | (-1.03) |
| HP | | | -0.0081 | (-0.65) | | | 0.0025 | (0.32) |
| SKEW | | | -0.0138 | (-0.91) | | | 0.0187 | (3.99) |
| BMOM | | | -0.0102 | (-0.67) | | | -0.0068 | (-0.89) |
| RB | | | -0.0085 | (-0.68) | | | 0.0002 | (0.03) |
| LIQ | | | -0.0229 | (-1.62) | | | 0.0018 | (0.27) |
| Curve-M | | | 0.0525 | (0.30) | | | -0.0143 | (-0.14) |
| Adj-$R^2$ | 0.18% | | -0.09% | | -0.59% | | 1.58% | |
| | | | | | | | | |
| **Panel C: All-asset benchmarks** | | | | | | | | |
| Alpha | 0.0102 | (3.50) | | | 0.0020 | (1.60) | | |
| All-asset value | 0.0639 | (2.30) | | | -0.0039 | (-0.33) | | |
| All-asset momentum | 0.0699 | (3.52) | | | 0.0107 | (1.28) | | |
| All-asset carry | -0.0144 | (-0.64) | | | 0.0065 | (0.67) | | |
| Adj-$R^2$ | 2.97% | | | | 0.33% | | | |

## Table 7. Risk versus Sentiment-Based Mispricing

The table tests whether the performance of the *S* strategy relates to business cycle risk or sentiment-based mispricing. Panel A summarizes the results from regressions of the *S* net excess returns onto business cycle variables. Panels B and C report the potential role of sentiment and investor psychology in explaining the net performance of the *S* strategy. TERM is the term spread, TED is the TED spread, DEF is the default spread, ΔFED is the change in the Fed funds rate, REAL is the 3-month real interest rate, LIQUID is the innovation in aggregate liquidity, ΔIP is the change in industrial production, CFNAI is the Chicago Fed National Activity Index, INFL is the inflation rate, UNC is the uncertainty index of Bekaert et al. (2022) multiplied by 100, $\beta(BW)$ is the sensitivity of the portfolio to the orthogonalized sentiment index of Baker and Wurgler (BW, 2006), High (H) and Low (L) BW are measured relative to the full-sample mean of the orthogonalized BW index. Mean, Alpha and Daily performance are annualized. Newey and West (1987) adjusted *t*-statistics are in parentheses. *p*-(H-L) denotes *p*-value for the null hypothesis: Alpha(High BW) = Alpha(Low BW). The sample covers the period from January 1992 to June 2019.

| | | Long | | Short | | Long - Short | |
|---|---|---|---|---|---|---|---|
| **Panel A: Risk-based explanation** | | | | | | | |
| TERM | | 0.1507 | (2.31) | -0.0545 | (-1.17) | 0.1026 | (2.32) |
| TED | | -0.1938 | (-1.90) | -0.1366 | (-1.31) | -0.0281 | (-0.35) |
| DEF | | 0.0011 | (0.86) | 0.0030 | (2.26) | -0.0010 | (-1.08) |
| ΔFED | | -0.0003 | (-0.15) | 0.0014 | (0.79) | -0.0008 | (-0.64) |
| REAL | | 0.1339 | (3.48) | -0.0087 | (-0.27) | 0.0713 | (2.70) |
| LIQUID | | 0.0047 | (0.80) | 0.0065 | (1.24) | -0.0008 | (-0.22) |
| ΔIP | | -0.0039 | (-0.41) | -0.0040 | (-0.43) | 0.0000 | (0.00) |
| CFNAI | | -0.0007 | (-0.80) | 0.0001 | (0.12) | -0.0004 | (-0.58) |
| INFL | | 0.0000 | (-0.24) | 0.0001 | (1.23) | -0.0001 | (-0.88) |
| UNC | | -0.0722 | (-0.14) | -0.6269 | (-1.43) | 0.2770 | (0.88) |
| | | | | | | | |
| **Panel B: Sentiment-based explanation** | | | | | | | |
| $\beta(BW)$ | | 0.0011 | (2.27) | -0.0012 | (-1.86) | 0.0011 | (2.56) |
| | | | | | | | |
| Mean | High BW | 0.0180 | (2.02) | -0.0324 | (-5.53) | 0.0252 | (4.96) |
| | Low BW | 0.0052 | (1.03) | -0.0061 | (-1.27) | 0.0056 | (1.76) |
| | *t*-(H-L) | | (1.49) | | (-3.60) | | (3.58) |
| | | | | | | | |
| Alpha | High BW | 0.0164 | (2.00) | -0.0357 | (-5.99) | 0.0261 | (5.08) |
| | Low BW | 0.0041 | (0.83) | -0.0059 | (-1.23) | 0.0050 | (1.59) |
| | *p*-(H-L) | | (0.20) | | (0.00) | | (0.00) |
| | | | | | | | |
| **Panel C: Day-of-the-week performance** | | | | | | | |
| Monday | | -0.0302 | (-3.37) | -0.0493 | (-6.11) | 0.0096 | (1.78) |
| Tuesday | | -0.0193 | (-2.07) | -0.0323 | (-4.52) | 0.0066 | (1.21) |
| Wednesday | | 0.0173 | (1.96) | -0.0074 | (-0.99) | 0.0123 | (2.17) |
| Thursday | | 0.0432 | (5.13) | -0.0053 | (-0.70) | 0.0242 | (4.30) |
| Friday | | 0.0412 | (4.88) | 0.0143 | (2.01) | 0.0134 | (2.70) |

**Table 8. Timing the *S* Strategy**

The table studies the performance of the *S* strategy that is timed with respect to the cross-sectional volatility of the slope parameter as averaged over the past *d* days. Panel A presents summary statistics for the performance of the timed *S* strategy, while Panel B reports estimated coefficients and adjusted goodness-of-fit statistics (Adj-*R²*) from spanning regressions of the timed *S* strategy onto the original *S* strategy. Mean, Volatility, Downside volatility, CER (certainty equivalent return) and Alpha are annualized. Newey and West (1987) adjusted *t*-statistics are reported in parentheses. The sample covers the period from January 1992 to June 2019.

| | *d*=3 | *d*=5 | *d*=10 | *d*=15 | *d*=22 |
|---|---|---|---|---|---|
| **Panel A: Summary statistics** | | | | | |
| Mean | 0.0190 | 0.0191 | 0.0191 | 0.0188 | 0.0184 |
| | (7.66) | (7.73) | (7.76) | (7.70) | (7.61) |
| Volatility | 0.0126 | 0.0126 | 0.0126 | 0.0126 | 0.0126 |
| Downside volatility | 0.0086 | 0.0085 | 0.0085 | 0.0085 | 0.0086 |
| Sharpe ratio | 1.5079 | 1.5187 | 1.5183 | 1.4958 | 1.4667 |
| Sortino ratio | 2.2465 | 2.2701 | 2.2733 | 2.2350 | 2.1862 |
| Omega ratio | 1.3141 | 1.3164 | 1.3164 | 1.3109 | 1.3038 |
| Skewness | 0.3132 | 0.3417 | 0.3800 | 0.3602 | 0.3401 |
| Excess kurtosis | 6.3012 | 6.7584 | 7.5972 | 7.5157 | 7.3080 |
| 99% VaR (Cornish-Fisher) | 0.0029 | 0.0029 | 0.0031 | 0.0031 | 0.0031 |
| % of positive months | 0.5434 | 0.5437 | 0.5443 | 0.5448 | 0.5455 |
| Maximum drawdown | -0.0271 | -0.0267 | -0.0265 | -0.0261 | -0.0261 |
| CER | 0.0186 | 0.0188 | 0.0188 | 0.0185 | 0.0181 |
| **Panel B: Spanning regression** | | | | | |
| Alpha | 0.0027 | 0.0028 | 0.0027 | 0.0024 | 0.0021 |
| | (2.77) | (2.89) | (2.96) | (2.71) | (2.40) |
| Beta | 0.9198 | 0.9215 | 0.9249 | 0.9268 | 0.9293 |
| | (45.02) | (45.13) | (45.36) | (45.65) | (46.26) |
| Adj-*R²* | 0.8461 | 0.8494 | 0.8557 | 0.8598 | 0.8649 |

## Table 9. Alternative Specifications of the Nelson-Siegel Model

The table reports summary statistics for the performance of the *S* strategy based on alternative specifications of the NS model. The alternative *S* strategies are based on longer dated term structures of up to 6 or 12 contracts in Panel A, on a seasonality-adjusted NS model for nine (NINE) or all (ALL) commodities in Panel B, or on smoother sorting signals using a 3- or 5-day moving average (MA) in Panel C. Mean, Volatility, Downside volatility, and CER (certainty equivalent return) are annualized. Newey and West (1987) adjusted *t*-statistics are reported in parentheses. The sample covers the period from January 1992 to June 2019.

|  | Panel A: Longer term structure | | Panel B: Seasonality adjustment | | Panel C: Smooth signals | |
|---|---|---|---|---|---|---|
|  | 6 contracts | 12 contracts | NINE | ALL | MA=3 | MA=5 |
| Mean | 0.0181 | 0.0184 | 0.0188 | 0.0197 | 0.0121 | 0.0117 |
|  | (6.18) | (4.66) | (4.77) | (5.27) | (4.92) | (4.70) |
| Volatility | 0.0152 | 0.0192 | 0.0193 | 0.0194 | 0.0127 | 0.0128 |
| Downside volatility | 0.0097 | 0.0122 | 0.0123 | 0.0127 | 0.0082 | 0.0082 |
| Sharpe ratio | 1.1909 | 0.9578 | 0.9737 | 1.0148 | 0.9538 | 0.9152 |
| Sortino ratio | 1.8896 | 1.5360 | 1.5544 | 1.5733 | 1.4884 | 1.4458 |
| Omega ratio | 1.2178 | 1.1766 | 1.1801 | 1.1883 | 1.1738 | 1.1639 |
| Skewness | 0.0350 | 0.3365 | 0.2932 | 0.1162 | 0.0730 | 0.0512 |
| Excess kurtosis | 2.8280 | 5.4492 | 5.5393 | 6.0164 | 2.5694 | 2.4379 |
| 99% VaR (Cornish-Fisher) | 0.0030 | 0.0048 | 0.0048 | 0.0048 | 0.0025 | 0.0024 |
| % of positive months | 0.5322 | 0.5221 | 0.5280 | 0.5276 | 0.5220 | 0.5216 |
| Maximum drawdown | -0.0333 | -0.0510 | -0.0626 | -0.0587 | -0.0567 | -0.0514 |
| CER | 0.0176 | 0.0177 | 0.0181 | 0.0189 | 0.0117 | 0.0114 |

**Table 10. Performance of Alternative Slope Strategies**

The table studies the performance of slope strategies that do not rely on Nelson and Siegel (1987) to model the slope signal. $\Delta S_t$ is the daily change in the slope $S_t = F_t^1 - F_t^4$ at time $t$, $F_t^k$ is the futures contract price at time $t$ with location $k$ on the term structure, $\Delta PC_{2,t}$ is the change in the second principal component at time $t$, and $RY_t$ is the roll yield at time $t$. Panel A presents performance statistics for the excess returns of these alternative slope strategies. Panel B reports estimated coefficients and adjusted goodness-of-fit statistics (Adj-$R^2$) from spanning regressions of the excess returns of the NS-based $S$ strategy onto the excess returns of these alternative slope strategies. Mean, Volatility, Downside volatility, CER (certainty equivalent return) and Alpha are annualized. Newey and West (1987) adjusted $t$-statistics are reported in parentheses. The sample covers the period from January 1992 to June 2019.

| | $\Delta S_t$ | $\Delta PC_{2,t}$ | $RY_t = F_t^1/F_t^k - 1$ | | | |
| | | | $k=2$ | $k=3$ | $k=6$ | $k=12$ |
|---|---|---|---|---|---|---|
| **Panel A: Summary statistics** | | | | | | |
| Mean | 0.0154 | 0.0074 | 0.0306 | 0.0242 | 0.0309 | 0.0365 |
| | (6.52) | (3.13) | (2.79) | (2.25) | (2.74) | (2.05) |
| Volatility | 0.0118 | 0.0121 | 0.0605 | 0.0603 | 0.0613 | 0.1087 |
| Downside volatility | 0.0076 | 0.0077 | 0.0397 | 0.0383 | 0.0392 | 0.0787 |
| Sharpe ratio | 1.3051 | 0.6091 | 0.5063 | 0.4011 | 0.5044 | 0.3359 |
| Sortino ratio | 2.0485 | 0.9756 | 0.8304 | 0.6875 | 0.8500 | 0.5511 |
| Omega ratio | 1.2423 | 1.1056 | 1.0898 | 1.0714 | 1.0899 | 1.0675 |
| Skewness | 0.0588 | 0.1399 | -0.1554 | 0.0276 | -0.0773 | -0.3675 |
| Excess kurtosis | 3.1473 | 2.9624 | 1.1854 | 1.2842 | 1.1767 | 4.4238 |
| 99% VaR (Cornish-Fisher) | 0.0023 | 0.0025 | 0.0105 | 0.0100 | 0.0104 | 0.0248 |
| % of positive months | 0.6364 | 0.6012 | 0.5858 | 0.5764 | 0.5854 | 0.5928 |
| Maximum drawdown | -0.0268 | -0.0411 | -0.1391 | -0.1545 | -0.1401 | -0.2455 |
| CER | 0.0151 | 0.0071 | 0.0233 | 0.0169 | 0.0234 | 0.0127 |
| **Panel B: Spanning regression** | | | | | | |
| Alpha | 0.009 | 0.018 | 0.018 | 0.018 | 0.018 | 0.018 |
| | (4.55) | (7.33) | (7.30) | (7.35) | (7.31) | (7.31) |
| Beta | 0.626 | 0.069 | 0.000 | -0.003 | 0.000 | 0.000 |
| | (33.34) | (3.56) | (-0.05) | (-0.79) | (0.07) | (0.12) |
| Adj-$R^2$ | 35.25% | 0.45% | 0.00% | 0.01% | 0.00% | 0.00% |

**Figure 1**. **Liquidity along the Curve**

This figure illustrates the percentage of open interest (OI) of the $m$th nearest contract and cumulative open interest (COI) of the nearest $m$ ($m$=1, 2, …12) contract(s) relative to the total open interest of the curve.

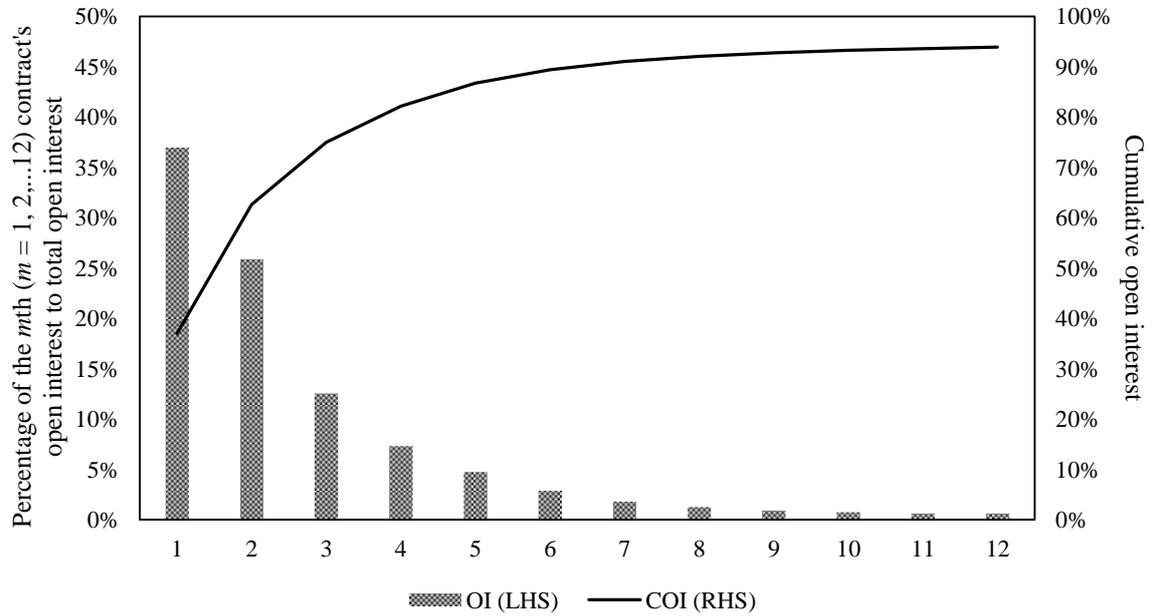

**Figure 2. Nelson-Siegel Parameters**

This figure illustrates the level betas and slope betas estimated from the NS model as well as the front futures prices (P1) of four commodities.

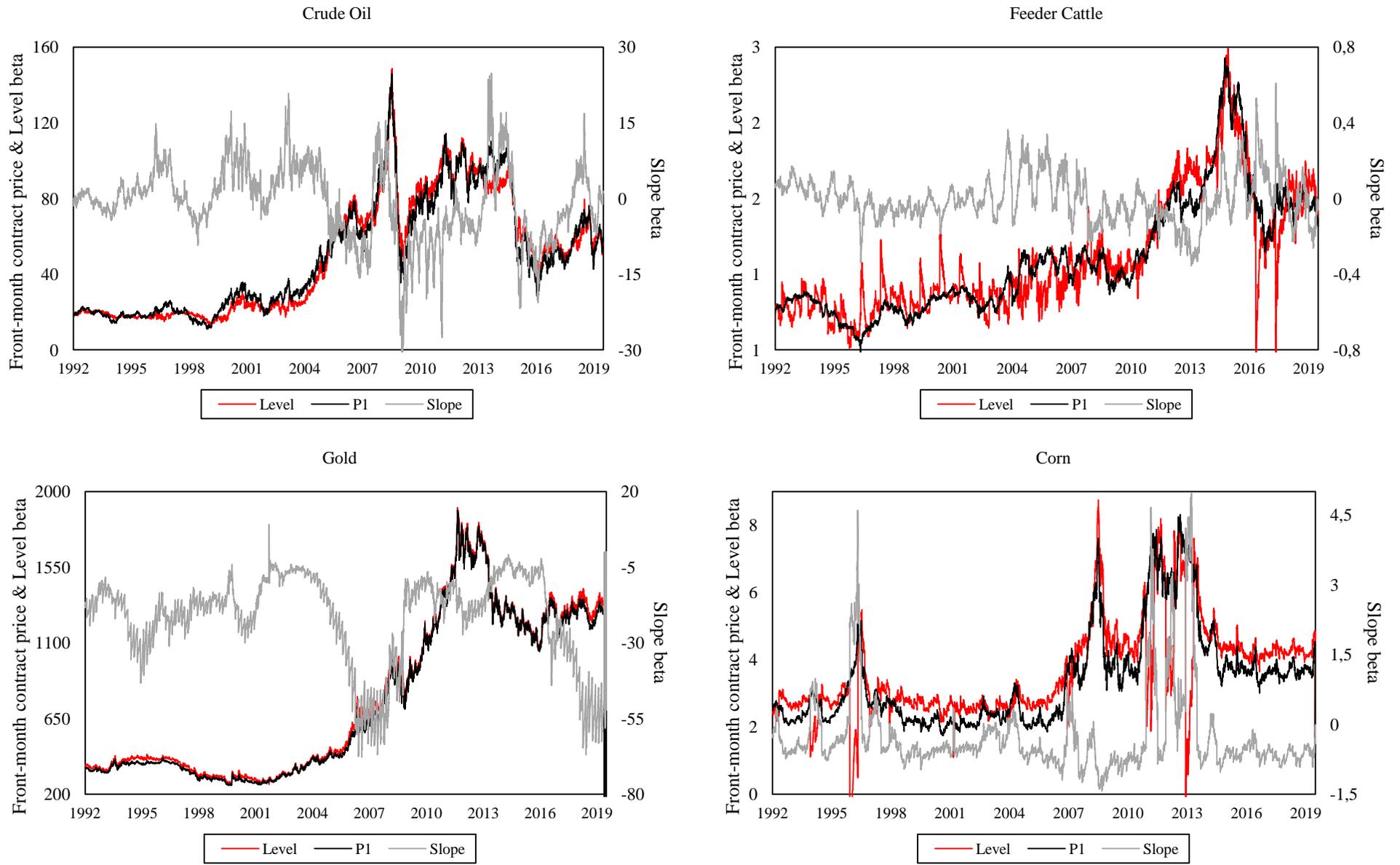

**Figure 3. Leverage and Value of $1 invested in the Timed-*S* Strategy**

This figure plots the value of $1 invested in the original and timed *S* strategies (left y-axis) as well as the leverage of the timed *S* strategy (right y-axis) over the period from January 1992 to June 2019. The wealth accumulated is based on total return which includes both futures excess return and collateral return.

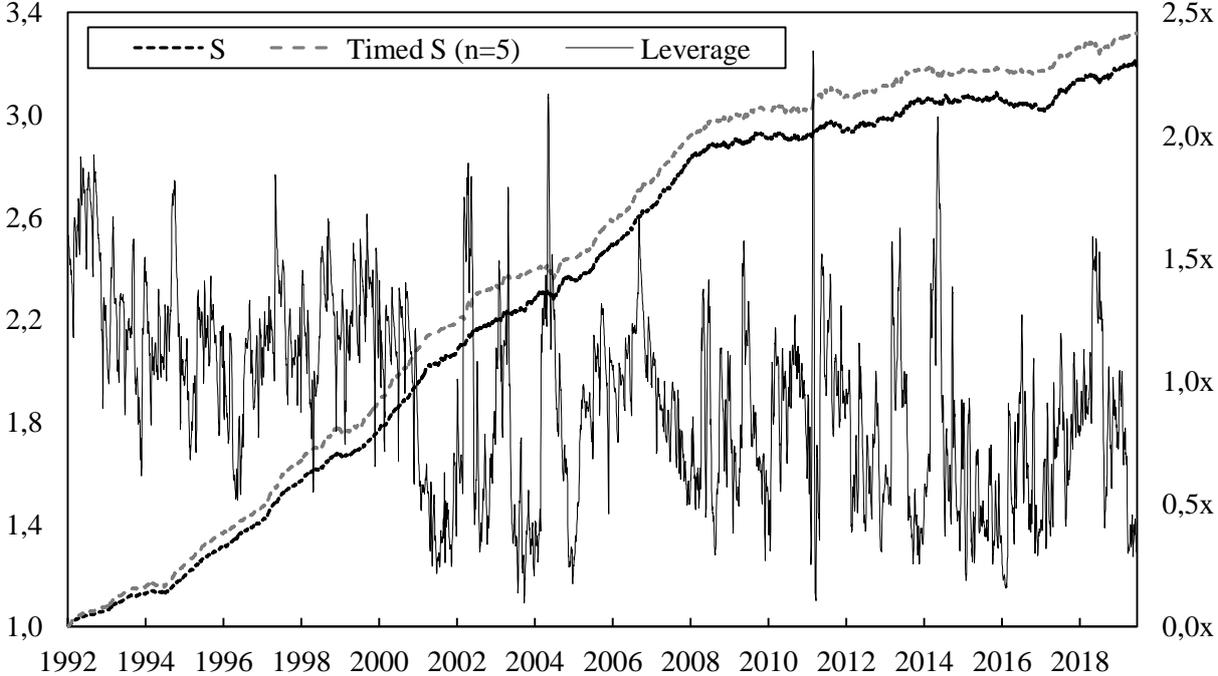

**Figure 4. Subsample and Sector Analyses**

Panel A illustrates the Sharpe ratios of the *S* portfolio over different subsample periods and of various commodity portfolios based on momentum (MOM), carry (CARRY), hedging pressure (HP), skewness (SKEW), basis-momentum (BMOM), relative basis (RB), and curve-momentum (Curve-M). Panel B presents the Sharpe ratios of sector-specific *S* strategies and the respective *t*-statistics (in parentheses) of the mean excess returns over the period from January 1992 to June 2019.

**Panel A: Subsample analysis**

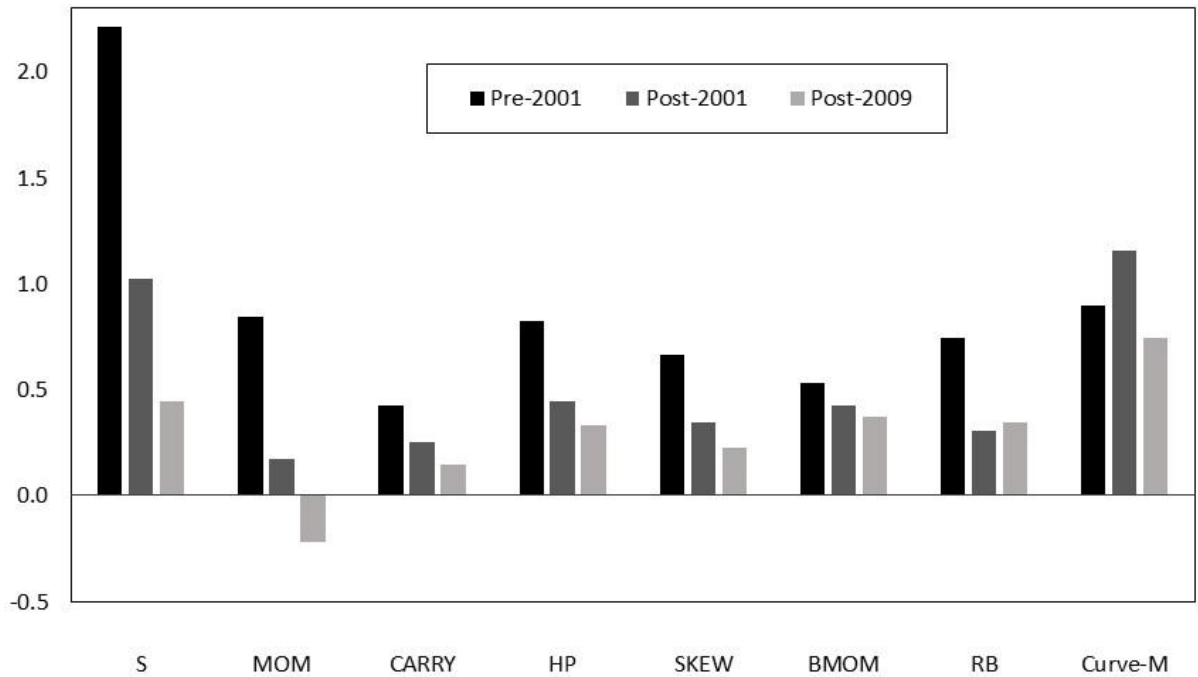

**Panel B: Sector analysis**

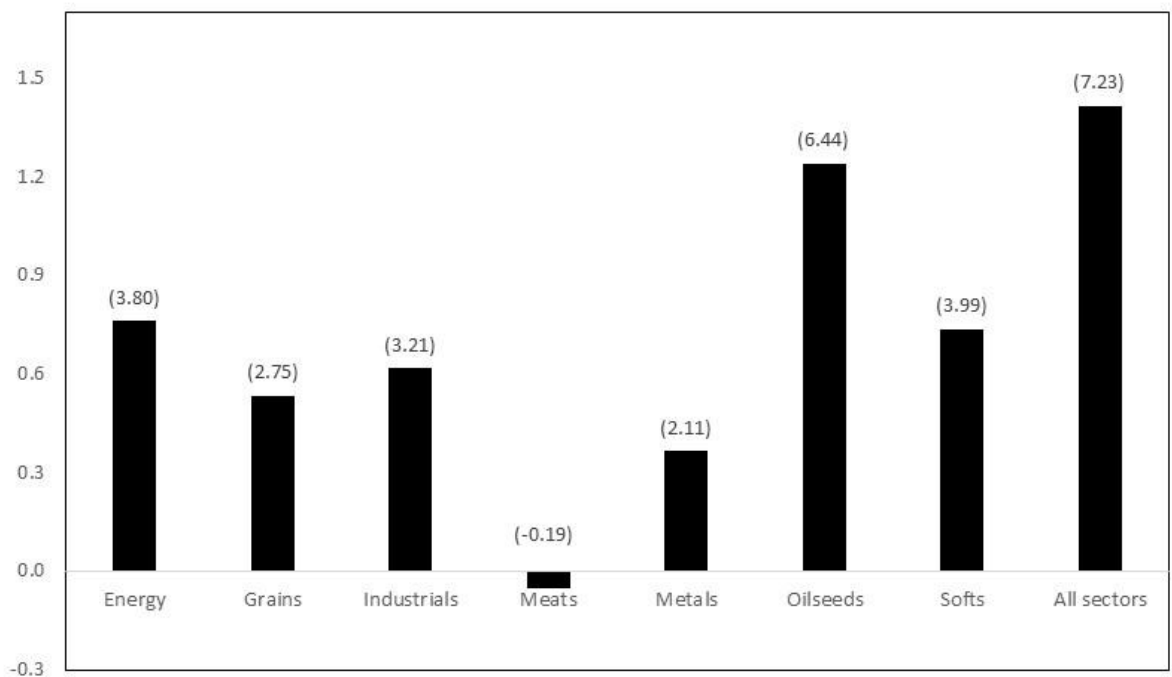